\documentclass[aps,twocolumn,superscriptaddress]{revtex4-2}
\pdfoutput=1
\usepackage{graphicx}
\usepackage{dcolumn}
\usepackage{bm}
\usepackage{amsmath}
\usepackage{amssymb}
\usepackage[pdfusetitle]{hyperref}
\usepackage[capitalize]{cleveref}
\crefname{equation}{Eq.}{Eqs.}
\Crefname{equation}{Equation}{Equations}
\crefrangeformat{equation}{Eqs.~(#3#1#4)-(#5#2#6)}

\usepackage{color}

\definecolor{nblue}{RGB}{28,130,185}
\definecolor{cgreen}{RGB}{76,153,0}
\definecolor{myorange}{RGB}{245,156,74}

\graphicspath{{Appendix}{Figures}}

\hypersetup{
  colorlinks=true,
  citecolor=cyan,
  urlcolor=cgreen
}

\def \beq{\begin{align}}
\def \eeq{\end{align}}

\newcommand{\br}{\bm{r}}
\newcommand{\bq}{\bm{q}}

\begin{document}

\title{Memory as activity: pattern formation in a conserved scalar field}
\author{Vaishnavi Gajendragad}
\affiliation{Max Planck Institute for Dynamics and Self-Organization (MPI-DS), D-37077 Göttingen, Germany}
\author{Suropriya Saha}
\email{suropriya.saha@ds.mpg.de}
\affiliation{Max Planck Institute for Dynamics and Self-Organization (MPI-DS), D-37077 Göttingen, Germany}

\date{\today}

\begin{abstract}
\noindent
We explore the concept of memory in scalar active matter, focusing on the collective dynamics of particles whose interactions depend on their evolutionary history rather than solely on their current configuration. We introduce the idea of an active particle whose velocity includes an active contribution that depends on its past trajectory suitably weighted by a memory kernel. The memory kernel is unrelated to the thermal noise acting on the particle, meaning that the dynamics breaks detailed balance at the microscopic level. We show that the number density of these active particles is described by a Cahn-Hilliard equation, which typically describes passive phase separation, suitably modified to account for this particular non-equilibrium effect. We establish the novel emergent features of the model and use the example of time delayed interactions to highlight its novel pattern-forming abilities through theory and simulations.
\end{abstract} 
\maketitle
\vspace{-1.2cm}
\section{Introduction} \label{sec:intro}
\noindent 
\begin{figure}
   \centering
  \includegraphics[width = \linewidth]{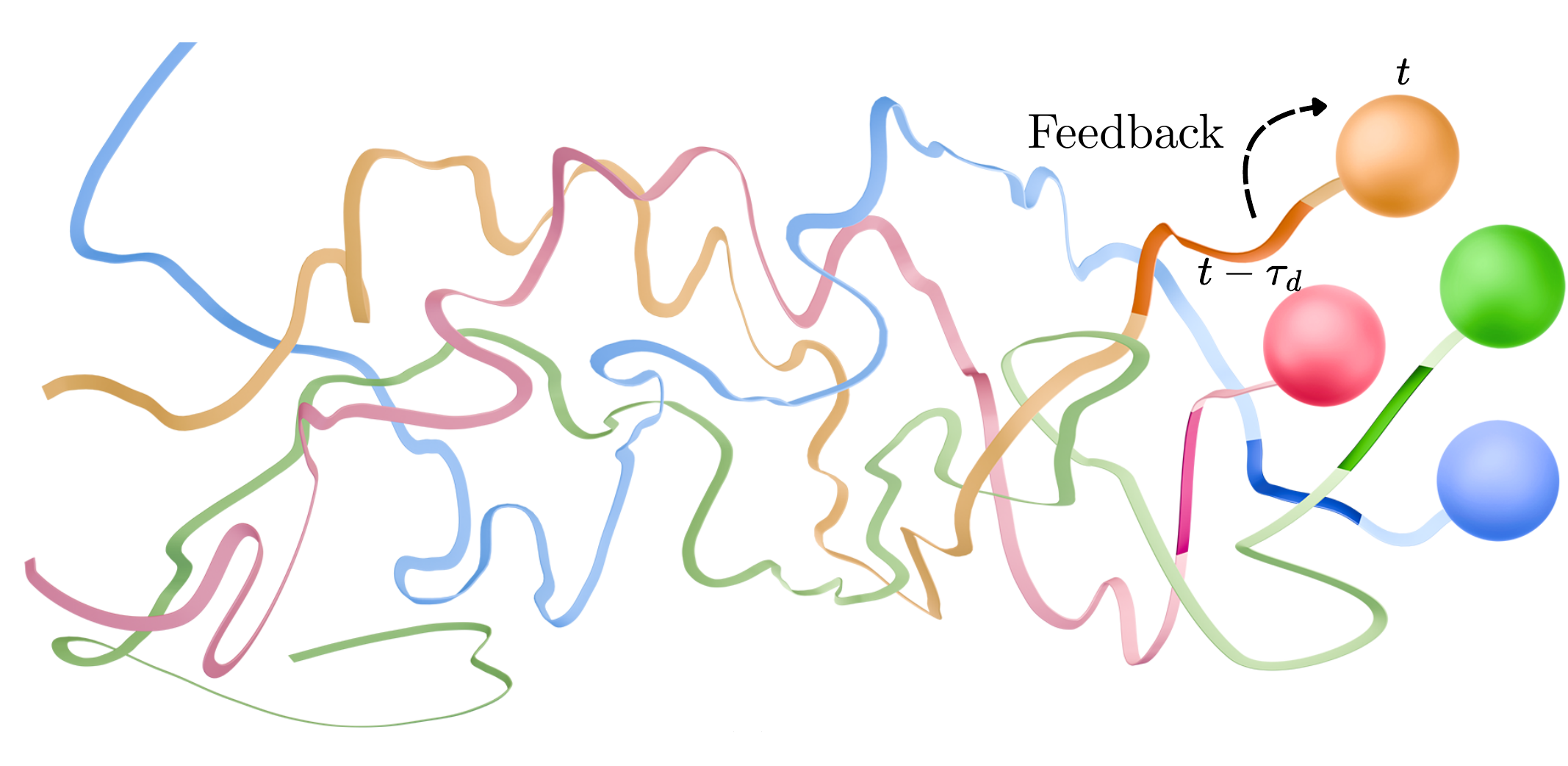}
  \caption{{\textbf{Memory as activity:}} Consider a scalar active particle that has access to its past trajectory (depicted by the coloured lines). The memory persists for a finite time, so the particle stores and utilizes information from a limited interval in the past (highlighted by the darker segment of the trajectory). The system is active due to the violation of the fluctuation–dissipation relation at the level of a single-particle. We explore the emergent dynamics of a collection of such particles, focusing first on the case where they receive feedback from a single instant $\tau_d$ in the past.}
  \label{fig:MemorySchematic1}
\end{figure}
\noindent 
We react to events and form opinions based on past experiences stored in our memory. The concept of memory, however, emerges naturally in systems far simpler than cognitive living beings with complex and adaptable neural networks \cite{Hopfield1982,bianchi2023self, miconi2017biologically, habashy2024adapting}. Supercooled liquids~\cite{Berthier_RevModPhys.83.587, scalliet2019rejuvenation} and shape-memory alloys~\cite{Song2013, stachiv2021shape} represent two classes of systems whose dynamical behaviour probed suitably reveals that they have developed memory. These systems retain signatures of their past evolutionary history which become evident in their response to externally applied fields or in their spontaneous fluctuations; see~\cite{RevModPhys.91.035002} for other systems across disciplines. Moving to the realm of active matter, consider the example of a self-propelled droplet driven by Marangoni flows: It both generates and responds to chemical trails left by others, thereby interacting with trajectories of the past and exhibiting caging and other signatures of glass-like behaviour~\cite{hokmabad2022chemotactic}.

Whether in systems relaxing toward equilibrium, or in those that are maintained away from equilibrium by energy injection at micro-scales, memory develops due to inherent nonlinearities, resulting in long-lasting dependence on initial conditions. As a consequence of these universal principles, systems as diverse as particles interacting via deposited chemical trails~\cite{Kranz2016TrailSelfInteraction,KRP_Saha_Golestanian_PhysRevLett.133.058401,dindo2024chemotactic,kumar2024emergent,dwivedi2025chemical}, tracers in supercooled liquids~\cite{Berthier_RevModPhys.83.587}, and dense tissues~\cite{Bray01032002} exhibit dynamics that bear striking similarities. Recently the concept has become relevant in the field of active matter, where reflecting the diversity of active systems, memory manifests due a myriad of reasons, for example non-reciprocity~\cite{Loos2020, Soto_PhysRevLett.112.068301,saha2019pairing}, collisions~\cite{caprini2024emergent}, and chemical or mechanical feedback mechanisms~\cite{ziepke2022multi, goth2025collective, yu2025feedback}.  

In this work we assume a different perspective and we will \emph{not} focus on the effects of memory that arise from complex many-body interactions of the physical system. The memory we consider is an intrinsic property of the active particle itself, see \cref{fig:MemorySchematic1}. Instead, using the toolbox developed in literature to investigate these systems~\cite{Zwanzig1973, Mori1965, zwanzig2001} to explore the collective behaviour of a collection of particles in which each individual can self-regulate using feedback from past interactions similar to crowds of human beings that constitute an example of information-mediated self-organisation~\cite{SocialForce_PhysRevE.51.4282}.  Our aim is to cater to scenarios where regulation processes are complex, perhaps driven by a combination of intelligence and information driven decision making, visual and verbal cues, or chemical signalling, such that it is impossible to model them exactly. 

\begin{figure}
   \centering
  \includegraphics[width = \linewidth]{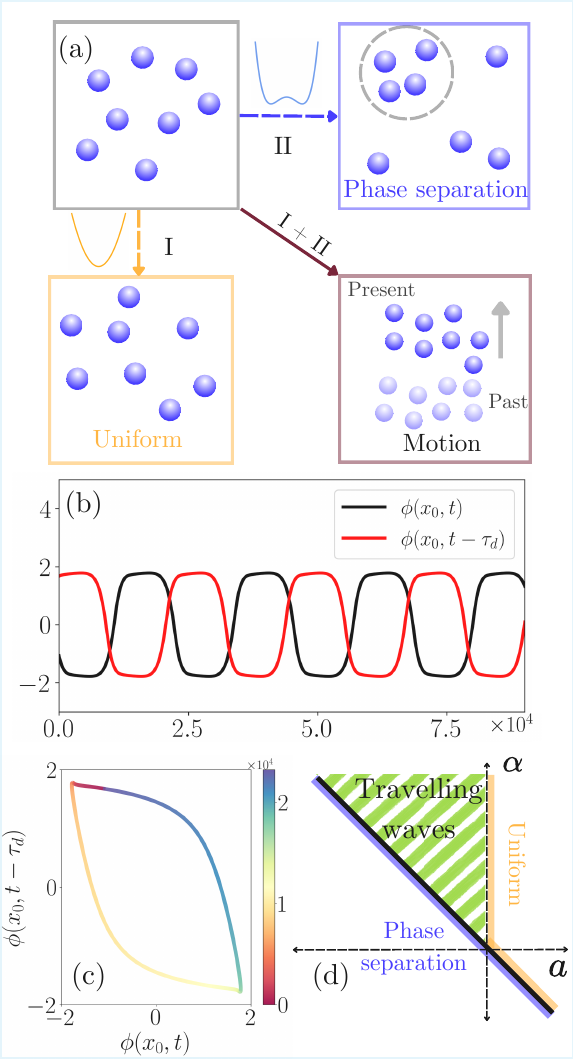}
  \caption{{\textbf{From active memory to moving patterns:}} (a) Travelling density patterns arise upon adding a history dependent contribution to phase-separation dynamics. The system evolves in two free energy landscapes with different mobilities: I has a parabolic free energy with frequency-dependent mobility, while II is a double well with constant mobility. (b) The juxtaposition of these contributions for a single species induces frustration, breaking time-reversal symmetry and producing oscillatory spatiotemporal dynamics, as evidenced by the time evolution of the density at a fixed position in one-dimensional simulations incorporating feedback from a time $\tau_d$ in the past, see \cref{fig:MemorySchematic1}. (c) Locally, at each lattice site a phase lag is established between the current and past densities which form Lissajous orbits when plotted against each other. (d) Two parameters—the feedback strength ($\alpha$) and effective inter-particle interaction ($a$) (see Sec.~\ref{sec:scalarMemory})—govern the qualitative behaviour. When they compete (top-left sector), spontaneously moving patterns emerge, including travelling waves, spirals, and, in more complex cases, irregular chaos.}
  \label{fig:MemorySchematic}
\end{figure}
We will explore how memory incorporated in a theory of scalar active particles with number conservation leads to novel collective behaviour. We choose scalar matter as perhaps the simplest system where activity can be incorporated in different ways such as via self-propulsion~\cite{Ramaswamy2010,wittkowski2014scalar,cates2025active}, non-reciprocity~\cite{sahaPRX_2020, parkavousi2025enhanced, rana2024defect}, and growth~\cite{Lish2024}, to open the door to studying interesting collective behaviour in a continuum field-theoretic approach within a minimal setup (minimum number of conserved fields). Activity is introduced in the system by incorporating contributions to the particle current that are determined not only by the instantaneous density but also by the history of its evolution, suitably weighted by a memory kernel. In this work, we consider time-translation invariant memory kernels, which are equivalent to frequency-dependent mobilities, and we therefore use these terms interchangeably. A simple example of our general model is a collection of particles, each of which receives feedback from a time $\tau_d$ in the past, see \cref{fig:MemorySchematic1}. The term representing what we call `active memory' supplements the thermodynamic fluxes that drive phase separation leading to stable bulk phase coexistence in the absence of activity.

 An interacting system governed by a free energy will eventually reach a stationary state that is a global minima of the free energy even if the mobility is frequency dependent~\cite{koide2006incorporating}, as illustrated in \cref{fig:MemorySchematic}(a) with two examples I and II. Our aim of constructing a system that operates out of equilibrium is achieved when the two contributions are superposed in the equation of motion of the density of a single species, meaning that the system evolves in two free energy landscapes. The dynamics is then no longer variational, as the frequency-dependent mobility in one contribution prevents relaxation to the minima of either free-energy landscape. This results in steady states that break time-reversal symmetry—an unambiguous signature of dynamics far from equilibrium. In fact, by introducing frustration where the system cannot minimise either free energy, we show that linearized dynamics involving suitable effective fields is non-Hermitian, drawing an analogy to non-reciprocal systems~\cite{saha2019pairing, Fruchart2021, sahaPRX_2020, You_Marchetti_2020}. 

Out of the various choices for the memory kernel, we focus on time-delayed interactions~\cite{loos2021stochastic,Loos2019,KoppKlapp2023} with applications in dynamical systems~\cite{Xu2018, SakamotoTakahiro2023}, and polar matter with aligning interactions~\cite{Zhou2024, Durve2018, TimeDelayViscek_PhysRevLett.127.258001,Holubec2021}. On incorporating time-delayed interactions in the continnum description of the density field, we find that it self-organises into a periodic structure in space such that there is a constant phase difference between its current configuration and the retarded configuration as seen in \cref{fig:MemorySchematic}(b). In effect, the past configuration of the field chases the current configuration such that a Lissajous orbit emerges in the steady state~\cite{strogatz2001nonlinear} as seen in \cref{fig:MemorySchematic}(c). The travelling waves are found when the strength of self-attraction parametrised by $a$, and the active contribution parametrised by $\alpha$, act antagonistically with one another in the region indicated in the $a$ -- $\alpha$ plane in \cref{fig:MemorySchematic}(d).

The paper is organized as follows. We begin with an introduction to the theoretical framework, Phase separation with memory, to describe the dynamics of a scalar active density in \cref{sec:scalarMemory}. We explain why this model represents a novel type of active phase separation, leading to pattern formation away from equilibrium. In \cref{sec:CH-delay}, we explore a particular example of the general model, namely, Cahn–Hilliard with time-delayed interactions,  where the system receives feedback from a time in the past. In \cref{Sec:linStable}, we perform a linear stability analysis around a homogeneous state to track the route taken by the system toward a patterning instability and discuss the role played by complex eigenvalues in determining the steady-state dynamics. By restricting to a simplified description of the system that retains just a single Fourier mode in space, we derive a relation between the angular frequency and wavelength of the travelling wave in \cref{sec:Dispersion}. We discuss wavenumber selection over a broad range of activity parameters for both deterministic evolution and in the presence of thermal fluctuations. Furthermore, we present an additional indicator of the nonequilibrium nature of the travelling wave solution by calculating the entropy production in the system and relating it to the phase difference that leads to the Lissajous figure in \cref{fig:MemorySchematic}(c). In \cref{sec:General}, we show that simple generalizations of the model lead to exotic dynamical patterns due to the same linear instability of the uniform state analysed in \cref{Sec:linStable}. In \cref{sec:Microscopics}, we provide an explicit microscopic motivation and coarse-graining procedure for the continuum model and we conclude with some final remarks in \cref{sec:conclusions}.

\noindent 
\section{Phase separation with memory} \label{sec:scalarMemory}
\noindent The conserved number density field $\phi(\br, t)$ evolves following the gradient of a chemical potential $\mu(\br, t)$ as follows
\begin{align}
    \partial_t \phi &= \Gamma \nabla^2 \mu + \sqrt{2 D} \bm{\nabla} \cdot \bm{\zeta}, \label{eq:continuity}
\end{align}
where $\Gamma$ denotes the mobility, $D$ is the diffusivity, and $\bm{\zeta}$ is a white noise term (delta correlated in space and time) with unit variance. We include a passive and an active contribution in $\mu$ writing it as 
\begin{align} \label{eq:MuCon}
    \mu \equiv \mu_{\rm eq} + \mu_{\rm ac}. 
\end{align}
The two parts of $\mu$ assume a similar structure
\begin{align} \label{eq:activeJ}
&\mu_{\rm eq}(\bm r, t) =  \int_{-\infty}^t \, \mbox{d} t' \gamma_{\rm eq}(t,t') \frac{\delta F}{\delta \phi}, \nonumber \\
&\mu_{\rm ac}(\bm r, t) = \alpha \int_{-\infty}^t \, \mbox{d} t' \gamma_{\rm ac}(t,t') \frac{\delta \bar{F}}{\delta \phi}.
\end{align} 

\begin{figure*}
    \centering
    \includegraphics[width=\linewidth]{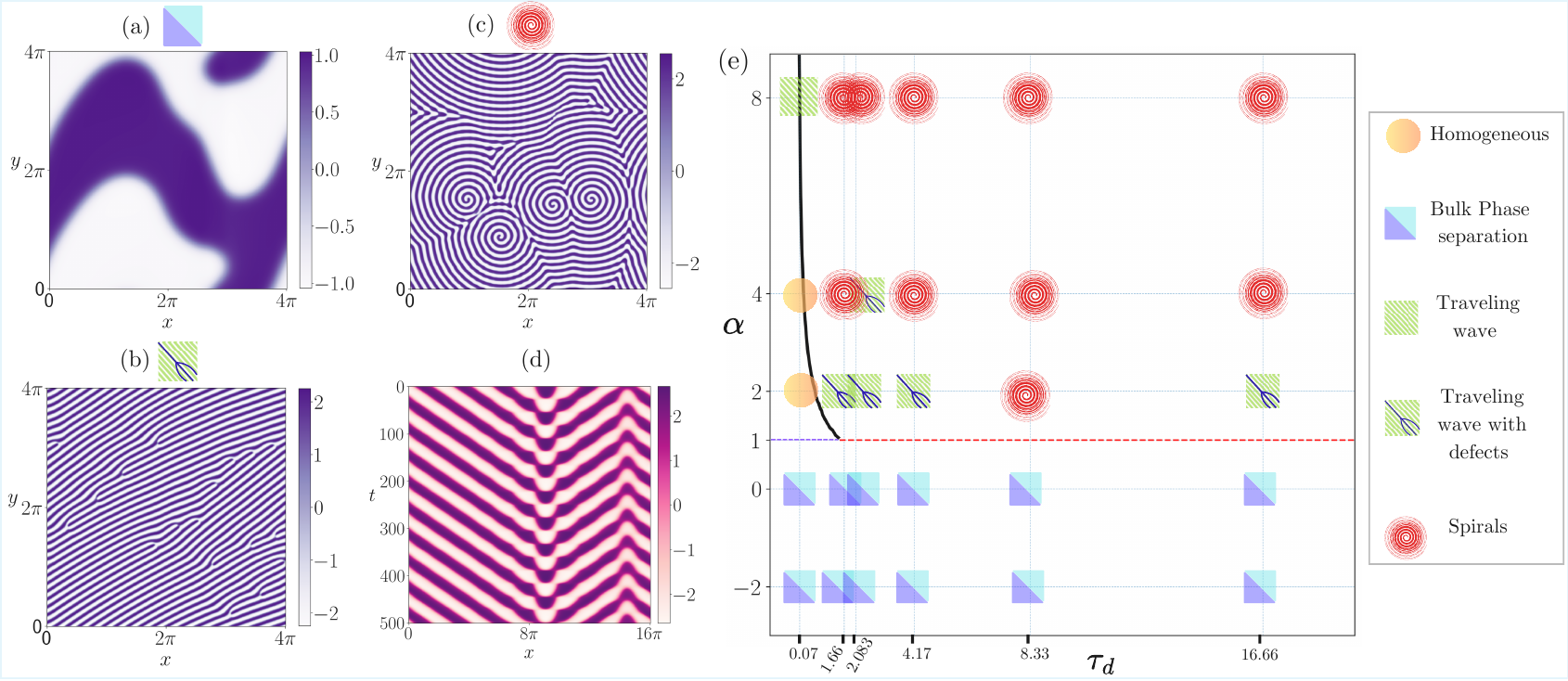}
    \caption{\textbf{Dynamical steady states and state diagram in Cahn-Hilliard with delay}: To illustrate the emergent collective behaviour possible within the general framework, we consider an example in which particles receive feedback from a time $\tau_d$ in the past with strength $\alpha$, a parameter that controls the level of activity in the system (see Eq.~\ref{eq:activeJ} and Sec.~\ref{sec:scalarMemory}). Exploring the phase space in the $\alpha-\tau_d$ plane in one and two spatial dimensions reveals four distinct stable steady states: a uniform state, bulk-phase separation, travelling waves (with or without long-lived defects), and spiral patterns, as shown in panels (a–d), see also MovieS1-S3 in S.M.~\cite{SI}. Phase separation into bulk domains is observed for $\alpha<0$ (see panel (c) of \cref{fig:MemorySchematic}), while travelling patterns arise when passive interactions promote attraction but the time-delayed interactions are repulsive; these patterns can often be accompanied by long-lived dislocations. At sufficiently large $\alpha$, multiple stable spirals are observed in the steady state over large parts of the phase diagram. Qualitatively similar features also appear in one dimension, as seen in the kymograph, where positionally ordered defects often persist in the steady state. The 2D phase diagram across varying $\alpha$ and $\tau_d$ summarizes these four regimes, and the transition from the uniform state to travelling patterns is predicted by a linear stability analysis of a homogeneous reference state. Linear stability analysis reveals transitions in the $\alpha, \tau_d$ plane. In (e) the black line marks the onset of finite-q instability, while the red line denotes instability at q=0, corresponding to bulk phase separation. The region between the black and blue lines represents homogeneous steady states where eigen-modes remain stable.}
    \label{fig:DynStates}
\end{figure*}

Instead of depending on the current configuration, $\mu$ depends on the history of evolution of the field weighed with a memory kernel $\gamma_{\rm eq,\rm ac}$. Memory kernels appear in a theoretical description when one focuses on a few selected degrees of freedom while eliminating others in a system with many-body interactions, as shown in the seminal work of Zwanzig and Mori~\cite{Zwanzig1973,Mori1965}, a formalism subsequently adapted to viscoelasticity~\cite{doi2013soft} and to mode-coupling approaches for glassy dynamics~\cite{10.3389/fphy.2018.00097}. In this paper we adapt the toolbox used to deal with systems in a complex medium and apply it to a system that receives feedback from its past. Here we do not provide any particular interpretation of $\gamma_{\rm eq,\rm ac}$, allowing it to be a property of the active system itself, as will be clear in Section~\ref{sec:Microscopics} where we provide a microscopic realisation of the dynamics that we introduce here phenomenologically. The scalar fields $F,\bar{F}$ are functionals of the field $\phi$ and its derivatives.

We focus solely on forms of $\gamma$ that are time-translation invariant, i.e. $\gamma_{\rm ac, \rm eq}(t,t') = \gamma_{\rm ac, \rm eq}(t-t')$. It is instructive to re-write $\mu_{\rm ac}$ in the Fourier domain, where $\gamma_{\rm ac}$ assumes the form of a frequency ($\omega$) dependent mobility 
\begin{align}
\mu_{\rm ac}(\bm q, \omega)
&= \alpha \gamma_{\rm ac}(\omega) \bar{\mu} (\bm q, \omega),
\label{eq:activeJFourier}
\end{align}
where $\bar{\mu} \equiv {\delta \bar{F}}/{\delta \phi}$. The integral of $\gamma_{\rm ac}$ over time is unity, i.e. $\gamma_{\rm ac}(\omega = 0) = 1$ such that the parameter $\alpha$ unambiguously tunes the level of activity in the system. The dynamics is thus a superposition of two chemical potentials with two frequency-dependent mobilities.

The terms in \cref{eq:activeJ} represent active and passive contributions because fluctuations in the steady state would follow the Fluctuation-Dissipation Theorem (FDT)~\cite{mazenko2006nonequilibrium} for $\alpha= 0$ if the noise terms are correlated as 
\begin{align}
    \langle \zeta_i(\br,t) \zeta_j(\br,t') \rangle = \delta_{ij} \gamma_{\rm eq} (|t-t'|),
\end{align}
while for $\alpha \neq 0$, FDT is violated. In this paper we choose  $\gamma_{\rm eq} = \delta(t-t') $ and simplify notation by dropping the subscript `$\rm ac$' in the definition of $\gamma_{\rm ac}$.

To demonstrate that the linearised dynamics of fluctuations about a uniform state is out of equilibrium, we consider the evolution of small perturbations $\delta \phi$ around that state. Restricting to second order in spatial gradients we have the following equation for the dynamics of $\delta \phi$ in the Fourier domain with external field $h$
\begin{align}
    \left(i \omega + D_{\rm eff}q^2 + \alpha \gamma(\omega) q^2 \right) \delta \phi( \boldsymbol{q}, \omega) =  h(\bq,\omega)+i \bq \cdot \bm{\zeta}. \label{lin_phiqom}
\end{align}
Defining density correlation function and response functions $C$ and $R$ in the usual way, $C(\bm{q}, \omega) = \langle |\phi(\bm{q}, \omega)|^2 \rangle$, and $R(\boldsymbol{q}, \omega) = \text{Im}[ \delta \phi/{\delta h}]_{h \to 0}  $, we find that the ratio~\cite{Chaikin_Lubensky_1995}
\begin{align}
    \frac{\omega C(\boldsymbol{q}, \omega)}{2DR(\boldsymbol{q}, \omega)} = \frac{1}{1 + (\alpha q^2 \gamma'' /\omega)}.\label{eq:FDratio}
\end{align}
The equation represents a frequency dependent deviation from unity, where $\gamma'' = \text{Im}[\gamma(\omega)]$, see \cref{sec:MetFDT} for the derivation. The ratio deviates from unity at finite $q$ as it should in the system with number conservation where density fluctuations as $q \to 0$ are suppressed. At a fixed $q$, the deviation is most significant at frequencies comparable to $\alpha q^2 \gamma''$.  

We now elucidate the conditions for which the \crefrange{eq:MuCon}{eq:activeJ} describes active phase separation for deterministic evolution. Notice that for $\gamma(\omega) = 1$, the model is simply passive model B with $\mu = \mu_{\rm eq} + \alpha \bar{\mu}$. If the two free energies $F$ and $\bar{F}$ are exactly the same, then it is possible to rewrite the deterministic dynamics for $\phi$ as $ i \omega \phi = -q^2 \Gamma [1 + \alpha \gamma( \omega)] \mu_{\rm eq} (\bm q, \omega)$, which also describes non-Markovian relaxation to a steady state determined by $F (= \bar{F})$. When the two contributions compete against one another, which happens when (1) $\bar{F} \neq {F}$, and (2) $\gamma(t)$ is a non-trivial function of time, we expect to observe dynamical steady states that carry explicit signatures that the system is out of equilibrium. 

$\mu_{\rm eq}$ is chosen such that for $\alpha = 0$, that is in the absence of activity, $\phi$ separates into bulk phases driven by thermodynamic forces and the composition of the bulk phases is determined by a Maxwell construction on the free energy $F$. In this paper we pick the following polynomial form for the chemical potential $\mu_{\rm eq}$, 
\begin{align}
    F &= \int_V \mbox{d}^d r' \left( \frac{a}{2} \phi^2 + \frac{b}{4} \phi^4 + \frac{\kappa}{2} |\bm \nabla \phi|^2 \right), \,\, \nonumber \\
\mu_{\rm eq} &= a\phi + b\phi^3 - \kappa \nabla^2 \phi, \label{eq:Mu}
\end{align}
where $\kappa$ is the surface tension that ensures that the interfaces are well-defined. In the rest of the paper we have scaled space and space using $\sqrt{\kappa}$ and $\kappa \Gamma^{-1}$, thus setting their values to unity. We have chosen the double well potential such that a passive system ($\alpha  = 0$) phase separates into two phases with compositions $\pm 1$ for vanishing average composition, $\bar{\phi} = 0 $. We choose the simplest possible form for $\bar{F}$
\begin{align}
    \bar{F} &= \int_V \mbox{d}^d r' \frac12 \phi^2, \notag \\
\bar{\mu} &= \phi \label{eq:MuP}
\end{align}

\section{Cahn-Hilliard with time-delayed interactions} \label{sec:CH-delay}
\noindent We now explore pattern formation in a model, where $\gamma (t) = \delta(t-\tau_d)$~\footnote{We distinguish the model from those cases where the Markovian dynamics of two or more fields can be formally written as the non-Markovian dynamics of a single field. }. In this simple version of the model, the particles receive feedback from a time $\tau_d$ in their past, reflected in the equation of motion of $\phi$ that reads
\begin{align}
    \partial_t \phi(\bm{r},t) &=  \nabla^2 \left[ a\phi + b \phi^3 \right] - \kappa \nabla^4 \phi \nonumber \\
 &+ \alpha  \nabla^2 \phi(\bm{r},t - \tau_d). \label{eq:CahnHilliardDelay}
\end{align}
\cref{eq:CahnHilliardDelay} represents dynamics away from equilibrium if both $\alpha, \tau_d \neq 0$. The signs of $a$ and $\alpha$ determine whether the particles constituting the field interact via attractive or repulsive interactions with the present and past field configurations respectively. 

\cref{eq:CahnHilliardDelay} is solved in a square domain with periodic boundary conditions using pseudo-spectral methods (details in~\cref{sec:MetNumerics}). As summarised in \cref{fig:DynStates}(e), for large enough $\tau_d$, dynamical patterns in the steady state are observed only if $a<0$, such that the system undergoes phase separation for $\alpha = 0$. The parameters of the free energy in \cref{eq:Mu} are fixed at $a=-1$, $b=1$, and the patterns formed in the system are explored by varying the parameters $\alpha$ and $\tau_d$; see \cref{fig:DynStates} for a summary of the dynamical states. Macroscopic phase separation occurs for $\alpha<1$, meaning that in this parameter regime the steady states do not reflect the non-equilibrium nature of the dynamics. Travelling patterns appear for $\alpha > 1$ when the particles at the present time interact attractively, while they interact repulsively with a past version of themselves. Another useful perspective is that when the feedback reinforces the dynamics imposed by $\mu_{\rm eq}$, bulk phases are found. When the feedback from the past competes with the present, coarsening is arrested and moving patterns emerge.

On increasing $\tau_d$ continuously from zero, while holding $\alpha$ constant, travelling patterns emerge above a threshold value of $\tau_d$ marked in \cref{fig:DynStates}(e). Interestingly, at values of $\tau_d$ smaller than the threshold, the uniform state is stable. The part of the phase space bounded by the dashed red line and the black line in \cref{fig:DynStates}(e) is dominated by spirals - as the name suggests these are travelling waves emanating outward from a point, see Supplemental MovieS1~\cite{SI}. They are particularly ubiquitous at larger values of $\tau_d$ although their stability depends on initial conditions. For small $\alpha$ and $\tau_d$ (close to the black transition line), a general trend is observed in the dynamics: although multiple spirals may form initially, they soon grow too large for the finite domain and eventually transition into a travelling-wave state with or without long-lived dislocations (see Supplemental MovieS2 and MovieS3~\cite{SI}), see \cref{fig:Formation_spirals} for series of heat-maps showing changes in $\phi$ as spirals are formed and stabilise. 

\begin{figure}
    \centering
    \includegraphics[width=\linewidth]{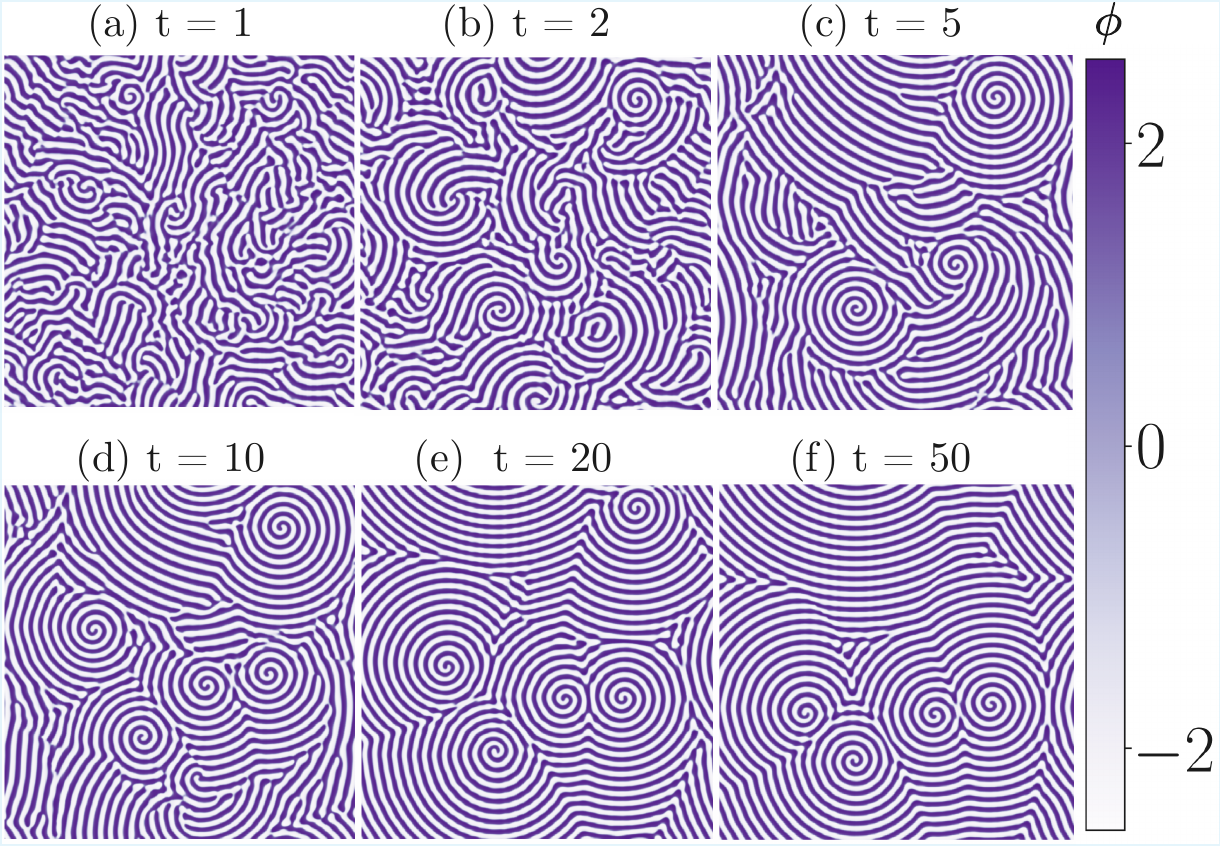}
    \caption{\textbf{Formation of spirals}: (a)-(c) Early stages in the formation of one or more spirals in the system. (d)-(f) depicts the later steady-state regime for $\alpha = 4.0, \tau_d = 8.33, \Delta t = 10^{-5}$. During the early stages, some spirals may grow to the system size by suppressing or eliminating others. In the steady state, multiple can spirals can coexist and maintain their stable indefinitely, see MovieS1 in S.M.~\cite{SI}.}
    \label{fig:Formation_spirals}
\end{figure}

\begin{figure*}
    \centering
    \includegraphics[width=\linewidth]{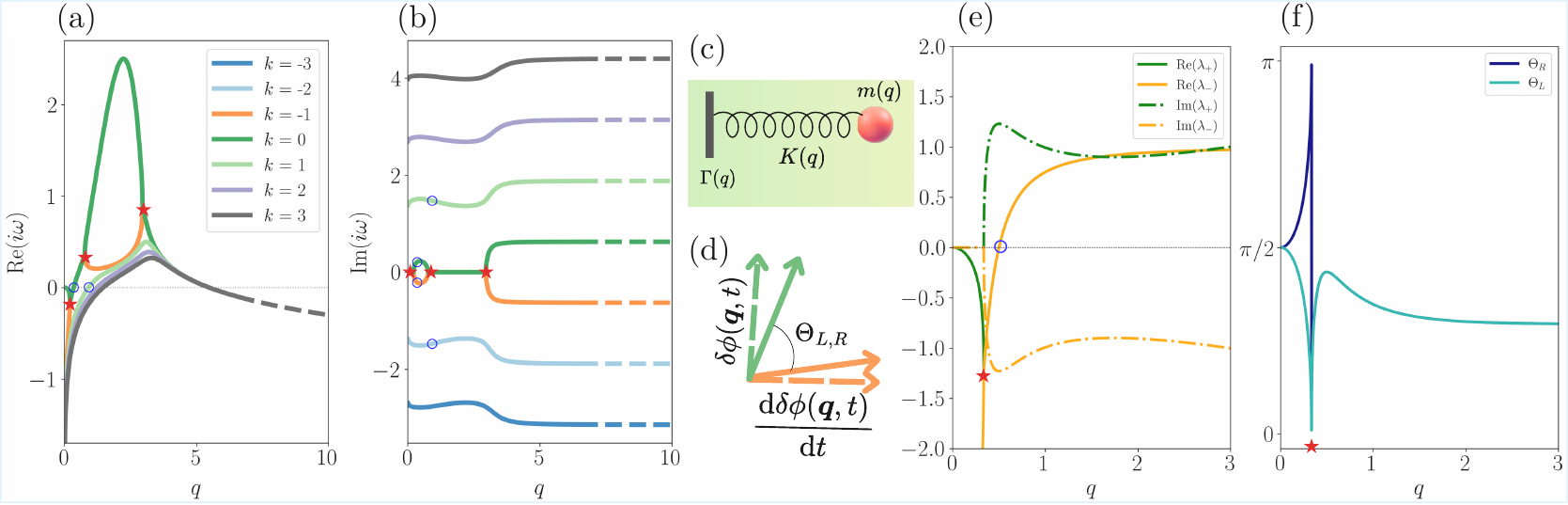}
    \caption{\textbf{Finite wavenumber linear instability of the uniform state} A linear stability analysis of the homogeneous state reveals multiple modes in the system whose real and imaginary parts are shown in panels (a) and (b) in different colors. The eigen-modes are branches of the Lambert-W function (labelled following the convention of the Lambert~$W$ function) and reflect the multivalued nature of the characteristic equation \cref{eq:lm_tra1}. (a) The principal branch $\lambda_{k=0}$ (green) approaches $0$ as $q \to 0$, while all others diverge to $-\infty$. Each branch becomes unstable at a finite value of $q$, with the corresponding critical $q_c$ shown as a function of $\tau_d$ in the Appendix. (a) and (b) shows three point where the $\mbox{Re}(i\omega)$ coalesce (marked with stars) suggesting that they are likely to be exceptional points (two roots coalesce and transition from real-complex or the reverse). The instability occurs via Hopf-bifurcation are denoted by black circle (a pair of complex eigenvalues change stability). Roots are computed using the Lambert~$W$ function (solid line) and the Newton–Raphson method (dashed line).  Parameters: $\alpha = 2, \tau_d = 5$. (c) A reduced description of the linearised dynamics shows that each Fourier mode mimics the dynamics of an inertial particle with wavenumber dependent mass, spring constant, and damping coefficients. The simplification shows that the EPs arise due to a coupling between the perturbations $\delta \phi$ and $v = \partial_t \delta \phi$. (e) Eigenvalues of the system (f) The eigenvectors are not orthogonal in this non-Hermitian system and become parallel at the EP.}
    \label{fig:LinearStability}
\end{figure*}

\section{Stability of the uniform state} \label{Sec:linStable}
\noindent
In this section, we probe the nature of the linear instability of the uniform state that leads to the emergence of the dynamic patterns. We perturb the homogeneous state allowing perturbations $\delta \phi$ around the constant value $\bar{\phi}$ and follow the time evolution of the perturbations. Substituting $\phi = \bar{\phi} + \delta \phi$ in \cref{eq:CahnHilliardDelay}, we obtain the linear equation for $\delta \phi$ in the Fourier domain as $\mathcal{L}(i\omega) \delta \phi(\bm{q},\omega) = 0$, where 
\begin{align}
    \mathcal{L}(i \omega) = i \omega- \left( {q}^2  - \kappa {q}^4  - \alpha {q}^2 e^{-i \omega \tau_d} \right). \label{eq:lm_tra1}
\end{align}
The signs of the real parts of the roots of $\mathcal{L}(i\omega)$, which can be complex, determine the stability of the uniform state. As the term proportional to $\alpha$ in \cref{eq:lm_tra1} is periodic in $\omega$, $\mathcal{L}$ has infinitely many roots at all finite values of $q$. The infinite spectrum of discrete roots of \cref{eq:lm_tra1} is evaluated analytically close to $q=0$ using the Lambert $W$ function (details in \cref{sec:MetLinStab}), and labelled using the terminology used to demarcate the branches of this special function \cite{kalugin2011analytical}. We adopt this terminology for ease of reference and mark the roots in \cref{fig:LinearStability} with distinct colours to denote the different branches. The roots are computed using a combination of analytic (solid line) and numerical methods (dashed line). 

In the rest of this section we will comment on the rich mode structure that we find, followed by a discussion on its physical significance. Constrained by number conservation, all roots coincide and become zero at $q=0$. As $q$ is increased, the $k=0$ branch changes continuously from zero, changes sign at a finite to signal an instability at a finite value of $q = q_c$. All other branches change discontinuously to non-zero values at finite $q$, and depending on parameters of the dynamics some of the lower branches change sign as well, always at wave-numbers larger that the $k=0$, which thus plays the most important role in determining the stability of the uniform state. The corresponding critical $q_c$ shown as a function of $\tau_d$ in the Methods \cref{Mfig:qc}. Other than $k=0,1$, all modes are always complex and appear in conjugate pairs. All branches eventually become indistinguishable at finite wavenumber and exhibit non-trivial decay behaviour, as detailed in \cref{Mfig:Branches}. 

The most important feature of the eigen-spectrum  is that the $k=0$ and $k=-1$ branches coalesce at an exceptional point (EP) at $q<q_c$, marked with stars in \cref{fig:LinearStability}, beyond which roots are complex. We find a direct correspondence between the values of $(\alpha,\tau_d)$ for which the roots follow the trend discussed here and the emergence of travelling patterns. The route to pattern formation is through a finite wave-number instability where a spectrum of wave-numbers are unstable. The structure of the eigenvalues are preserved for both signs of $\alpha$.

The analysis provides several important pieces of information. It shows infinitely many modes participate in the dynamics at small $q$, one of which vanishes at $q\rightarrow0$ respecting the number-conservation law. The oscillatory response arises due to the active term due to the participation of the two branches $k=0,-1$, which raises an important question - can we identify the eigenvectors corresponding to the two modes and thus identify the fields that participate in the dynamics leading to the complex eigenvalues -  a clear sign of non-hermiticity.

\subsection{Active memory acts as inertia} \label{SubSec:Inertia}
\noindent
To establish that the points at which the two branches labelled $k = 0$, and $k = -1$ coincide (see \cref{fig:LinearStability}(a)) is indeed an exceptional point, we consider a reduced description of the linearized dynamics. A Taylor series expansion of the active term in \cref{eq:CahnHilliardDelay} retaining terms until second order in $\tau_d^2$ yields the following equation of motion for the perturbations $\delta\phi_{\bm{q}}$ and generalised velocity $v \equiv\dot{\delta \phi}$
\begin{align}
&\partial_t \delta \phi = v, \nonumber \\
& m(q) \partial_t  v + \Gamma'(q) v   +  K(q) \delta \phi = i \bm{q} \cdot \bm{\zeta}(\bq,t)  \label{2nd_order_exp}
\end{align}
which resembles the dynamics of an inertial particle in a harmonic trap with the important distinction that all coefficients are wavenumber dependent, as depicted in \cref{fig:LinearStability}(c). The mass, damping, and spring-constant are given by 
\begin{align}
& \Gamma'(q) =  (1 - \alpha \tau_d q^2), \nonumber \\
& m(q) = \frac12 \alpha \tau_d^2 q^2, \nonumber \\
& K(q) = q^2( (a + \alpha)  + \kappa q^2 ). \label{eq:ReducedEq}
\end{align}
An important conclusion from the reduction is that the effect of the feedback is to confer inertia to a mass-less system. The coupled dynamics of $\delta \phi$ and $v$ leads to complex eigenvalues which are
\begin{align}
   & \lambda_{\pm} = -\frac{\Gamma'(q)}{2 m(q)} \pm \frac{\sqrt{\Gamma'(q)^2 - 4 m(q) K(q)}}{2 m(q)}  \nonumber \\
    &= -\frac{(1 - \alpha \tau_d q^2)}{\alpha \tau_d^2 q^2} \pm {\sqrt{\frac{(1 - \alpha \tau_d q^2)^2}{\alpha^2 \tau_d^4 q^4} - 2\frac{a + \alpha  + \kappa q^2}{\alpha \tau_d^2}}}
\end{align}
On calculating the left and right eigenvectors of the eigenvalues, we observe that the eigenvectors are not orthonormal to each other reflecting the  non-hermiticity of the system. Both sets of eigenvectors, left or right, coincide at the exceptional point where $\Theta_{L,R}$ vanish. A Hopf bifurcation point at a finite $q$ succeeds the EP at the point where the real parts of $\lambda_{\pm}$ vanish. Notice that beyond the point of Hopf-bifurcation, $\lambda_{\pm}$ are unstable. $\Gamma'(q)$ vanishes at a finite $q$ only for $\alpha>0$ at which point $\lambda_{\pm}$ is complex if $(a+\alpha) + \kappa q_c^2 > 0 $. Assuming that $\kappa$ is small, the condition is met for $a+\alpha>0$ which is the section marked in \cref{fig:MemorySchematic}(c).  

It is likely that in the full dynamics, the occurrence of multiple branches corresponds to higher order derivatives of $\delta \phi$. For increasing $\tau_d$ more and more terms have to be included in the stability analysis, meaning that the dynamics becomes sensitive to the whole history of evolution of the system.  

As shown in the S.M.~\cite{SI}, the entropy produced by fluctuations around the uniform state diverges meaning that the homogeneous state is unstable leading to the formation of patterns. See also the section with a discussion of the critical wavenumber $q_c$ above which the system is unstable. we find that $q_c$ decreases as $\alpha$ increases.

We summarise the route to non-hermiticity in the following way - the number density couples to the rate of change of density fluctuations leading to an EP at finite wavenumber. The real part of the pair of eigenvalues changes sign at a larger value of $q$ for signalling a Hopf-bifurcation which appears when the effective damping $\gamma$ vanishes.   
The very important difference that distinguishes the instability from others reported in literature is that it does not involve two separate fields.

\section{Dispersion relation and wavenumber selection} \label{sec:Dispersion}
\begin{figure}[h]
    \centering
    \includegraphics[width=0.8\linewidth]{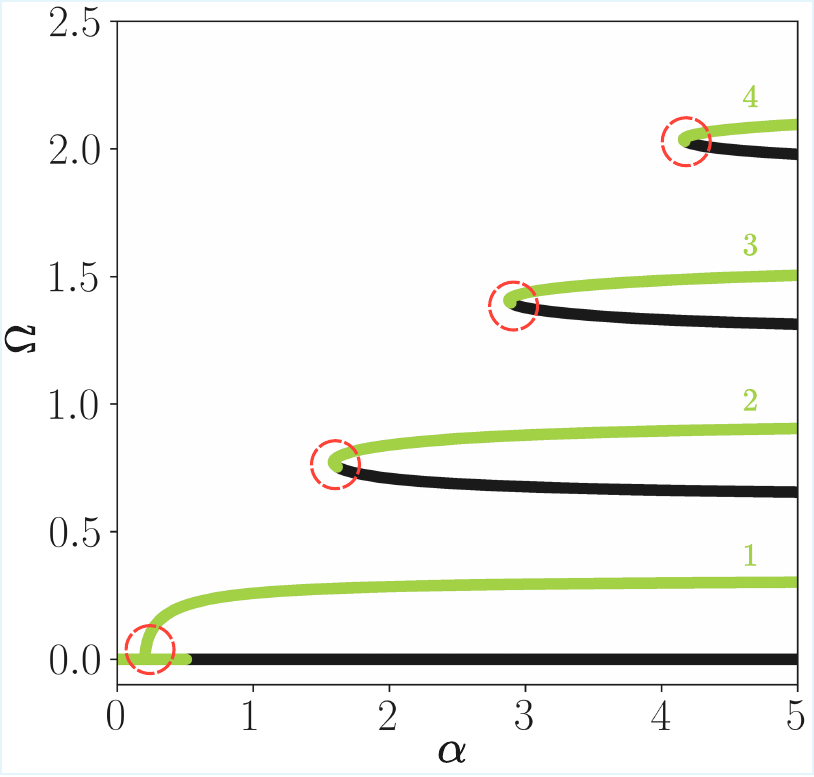}
  \caption{\textbf{Dispersion relation} Roots of \eqref{eq:DispRelation} obtained for a fixed $\tau_d = 10$, $q_0 = 0.7$. Green lines indicate physical solutions that satisfy the amplitude condition and black indicates roots that do not. Hence, as seen in the plot, the trivial solution ($\Omega = 0$) is not viable beyond $\alpha_c = 0.51$, marking the point at which the phase separated state is unstable. $\Omega$ is a multivalued function of $\alpha$, with the number of possible solutions increasing as $\alpha$ increases. The bifurcation points(indicated by red dotted circles), at which the number of solutions of the equation changes, see S.M.~\cite{SI} for details.}   
  \label{fig:Diespersion}
\end{figure}

\noindent 
In a system of size $L$, we truncate the Fourier representation of $\phi(x, t)$ to the fundamental wavenumber $q_0 = 2\pi/L$ and frequency $\Omega$ only to write it as: 
\begin{align}
    \phi = A\exp[{i(q_0 x - \Omega t) }] + A^*\exp[{-i (q_0 x - \omega t)}] \label{eq:TWansatz}
\end{align}
Substituting the ansatz in \cref{eq:TWansatz} into \cref{eq:CahnHilliardDelay}, and truncating the expansion to the single mode by simply ignoring all higher modes generated by the nonlinear terms, we get the following relation between the angular frequency $\Omega$ and the wavenumber $q_0$
\begin{align}
    \Omega &= \alpha q_0^2 \sin (\Omega \tau_d),
    \label{eq:DispRelation}
\end{align}
The amplitude satisfies
\begin{align} 
    A^2 &= \frac{1}{3} \bigg( 1 - \kappa q_0^2 - \alpha \cos(\Omega \tau_d) \bigg).
    \label{eq:Ampcond}
\end{align}
The amplitude equation \cref{eq:Ampcond} provides a condition for the solution to be physical. The transcendental equation \cref{eq:DispRelation} and solutions that satisfy the amplitude condition is plotted in \cref{fig:Diespersion}. 
The trivial solution $\Omega = 0$, the critical $\alpha_c$ at which the the root becomes unstable is given by:
\begin{equation}
    \alpha_c = 1 - \kappa q_0^2
\end{equation}
The theoretical predictions of the dispersion relation are verified by numerical simulations in \cref{fig:Diespersion}(d), where we observe excellent agreement between the theoretically and numercially obtained $\Omega$ despite the simplicity of the ansatz used in \cref{eq:TWansatz}.

\begin{figure}
    \centering
    \includegraphics[width=\linewidth]{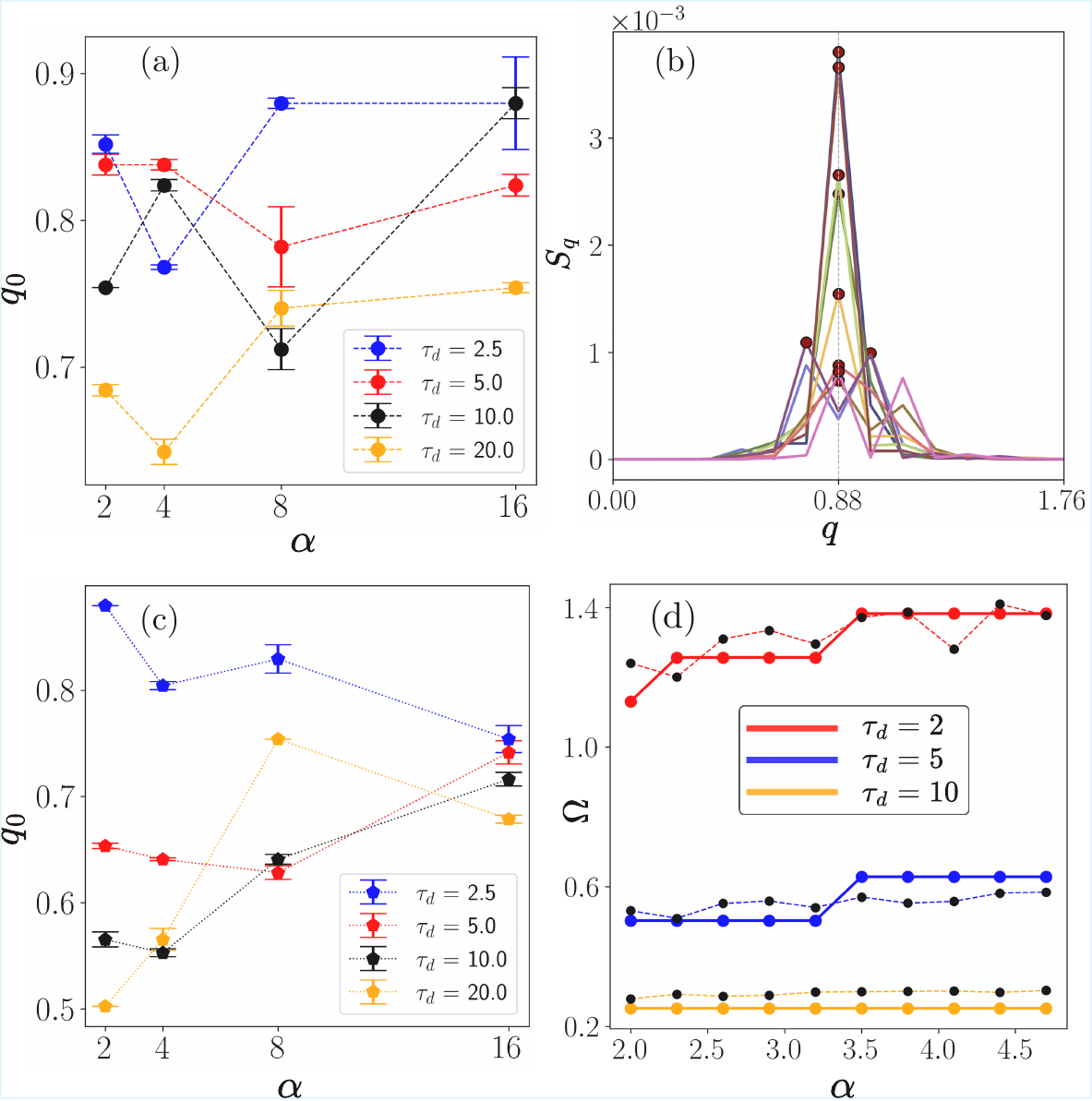}
    \caption{\textbf{Wavenumber selection and numerical verification of the dispersion relation in 1D.} (a) The wavenumber selection across a range of $\alpha$ and $\tau_d$ values, where each point represents the mean of 10 realizations and the error bars denote the variance across them. (b) The static structure factor for all realizations for $\alpha = 4$ and $\tau_d = 10$ peaks at $q = 0.88$.  (c) The same analysis as in (a) in the presence of noise ($D = 10^{-4}$). For $\alpha = 16$ and $\tau_d \in (2.5, 20)$, the average variance decreases from $\sigma^2 = 0.013256$ without noise to $\sigma^2 =  0.0084483$ with noise. (d) Numerical verification of dispersion the relation for for $\tau_d$ shown in the inset. The steady-state $\Omega$ across the considered $\alpha$ values exclusively fall on the first non-trivial branch. The black dots show the theoretical $\Omega_{Th}$. We observe that as $\tau_d$ increases the numerical agreement to the theoretical $\Omega_{Th}$ reduces. All panels correspond to a system of size $N = 512$ and $L = 50$, additional results on finite size effects are included the S.M.~\cite{SI}.}
    \label{fig:WavenumberDispersion}
\end{figure}
 
The system exhibits strong wavenumber selection in the steady state, indicating that system possess a intrinsic length scale determined by the activity parameters $\alpha, \tau_d$ independent of initial conditions. For each pair of parameters $\alpha$ and $\tau_d$, we identify the selected wavenumber $q_0$ corresponding to the peak of the static structure factor, defined as
\begin{align}
C(q) = |\phi(q)|^2 .
\label{eq:Sq}
\end{align}
To eliminate the dependence on initial conditions, we compute the mean and variance of the selected wavenumber $q_0$ over 10 unique realizations. As shown in \cref{fig:WavenumberDispersion}(a, c), the ensemble variance of $q_0$ is small, indicating robust wavenumber selection. Notably, in the presence of noise, the average variance across the parameter space $\alpha$ in the range $2-16$ and $\tau_d$ in the range $2.5-20$ decreases by 64\%, demonstrating enhanced selection robustness. This behaviour may arise from stochastic fluctuations that suppress competing modes and enhance convergence to the dominant mode, although a detailed analysis of this mechanism remains for future work.
This consistently low variance across the explored parameter space indicates that wavelength selection is a robust and reproducible feature of the dynamics, for further details see S.M.~\cite{SI}.

Using the ansatz in \cref{eq:TWansatz} we can calculate the local entropy production rate in the steady state which is given by the expression $S = -\langle \mu_{\rm ac} \dot{\phi} \rangle D^{-1}$ \cite{nardini2017entropy, alston2023irreversibility, suchanek2023entropy} to further provide insight into the nonequilibrium nature of the model. Substituting the travelling wave solution into the expression for S, we obtain an expression for the mean entropy produced per unit area and time,
\begin{align}
    S = \frac{\alpha |A|^2}{2D} \Omega \sin(\Omega \tau_d). \label{eq:entropy}
\end{align} 
$S$ is non-zero only when $\tau_d \Omega  \neq 0$, meaning that the system is active only when a feedback mechanism leads to travelling patterns with angular frequency $\Omega$. The expression in \cref{eq:entropy} is verified numerically by calculating the $S$ in the steady state. For a given $\alpha$ and $\tau_d$, $|A|$ is constant due to wavenumber selection, and $\Omega$ becomes independent of $\alpha$ for large enough values, see \cref{fig:Diespersion} and \cref{fig:WavenumberDispersion}, suggesting that $S$ increases linearly with activity, see S.M.~\cite{SI} for details.

\section{Generalizations of the model} \label{sec:General}
\noindent 
Cahn-Hilliard with delay represents a simple realization of the general model proposed in Sec.~\ref{sec:scalarMemory}. We show using two generalizations where we include a second time-scale in the system by introducing another term that provides feedback to the system. The aim is to illustrate that when $\gamma$ and $\bar{\mu}$ are allowed to assume other forms, several new features emerge in the dynamics. 
\subsection{Route to the moving patterns}
\noindent Consider the linearised description with a general memory kernel $\gamma$
\begin{align}\label{eq:GenMem}
    i\omega = -q^2(a + \kappa q^2) - \alpha q^2\gamma(i\omega)
\end{align}
\cref{eq:GenMem} is in general a transcendental equation can have many solutions at a  given $q$.  Taylor expanding $\gamma(i\omega)$ as $\gamma(i \omega) = \dot{\gamma}(0) i \omega - \omega^2 \ddot{\gamma}(0) $ and substituting this in \cref{eq:GenMem}, we get the following quadratic equation for the reduced description 
\begin{align}
    i\omega\Gamma'(q) + m(q)(i\omega)^2 &= K(q),
\end{align}
where 
\begin{align}
    \Gamma'(q) &= 1 - \alpha q^2 \dot{\gamma}(0) \notag \\
    m(q) &= \frac{\alpha q^2 \ddot{\gamma}(0)}{2} \notag \\
    K(q) &= -q^2(a + \kappa q^2 +\alpha \gamma(0))
\end{align}
When the memory kernel is of the form of a single time delay $\gamma(\omega) = \exp(i\omega \tau_d)$, the constants have the form given in Eq.~\eqref{eq:ReducedEq}. When there are multiple memory kernels of the form $\gamma(\omega) = \exp(i\omega \tau_{d, 1}) + \exp(i\omega \tau_{d, 2}) $, the constants have the following form:
\begin{align}
    \Gamma'(q) &= 1 - \alpha q^2(\tau_{d, 1}  + \tau_{d, 2}) \notag \\
    m(q) &= \frac{\alpha q^2 (\tau_{d, 1}^2  + \tau_{d, 2}^2)}{2} \notag \\
    K(q) &= -q^2(a + \kappa q^2 +\alpha )
\end{align}
The eigenvalues of the reduced model describe the dynamics of the two most important modes (see Fig. M6 of S.M.~\cite{SI} ). This analysis generalises the conclusions of \cref{SubSec:Inertia} to any form of memory that admits a series expansion in $\omega$, showing that travelling patterns arise due to a non-Hermitian coupling between $\delta \phi$ and $\partial_t \delta \phi$.
\subsection{Steady state dynamics}
\begin{figure*}
    \centering
    \includegraphics[width=0.9\linewidth]{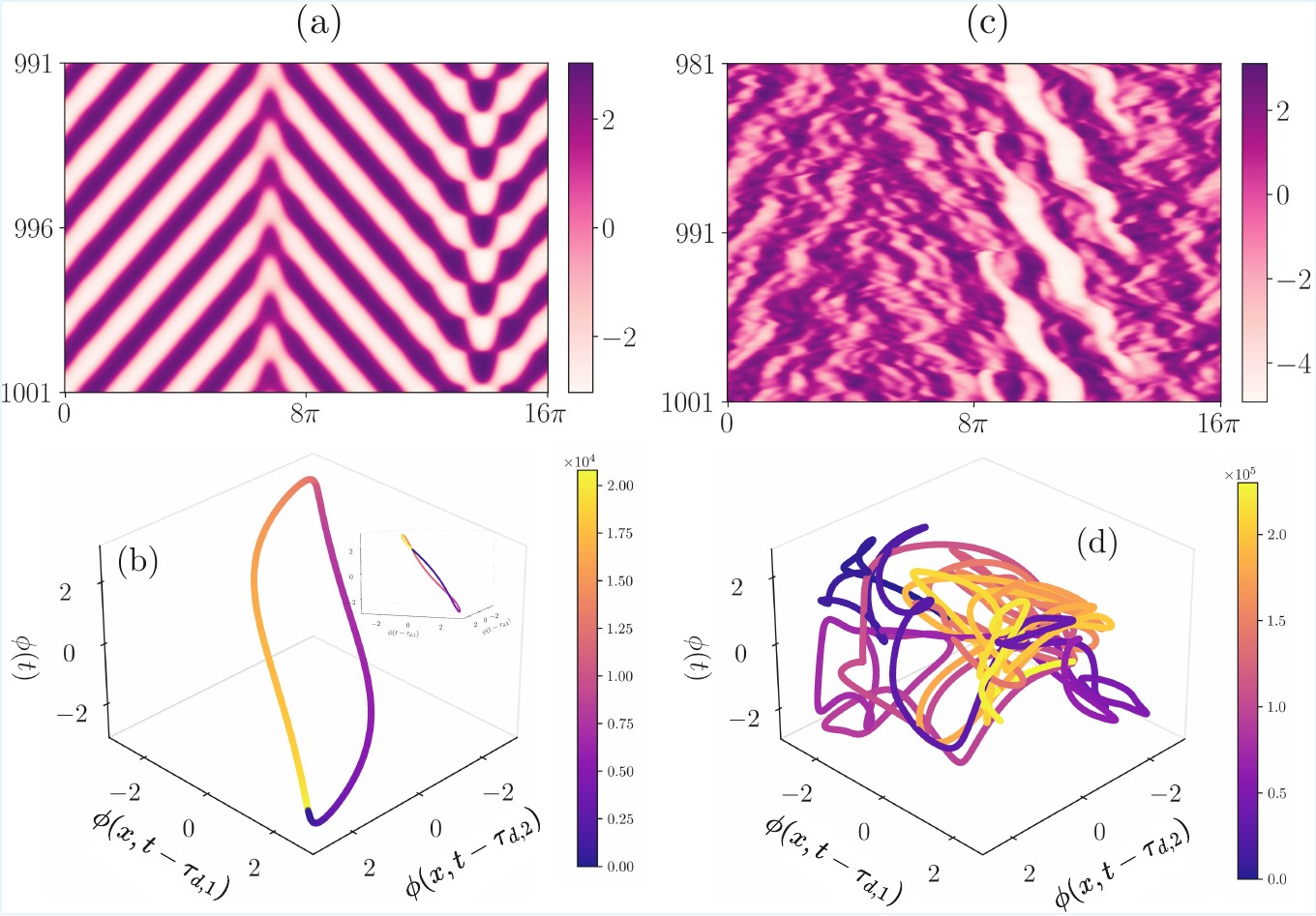}
  \caption{\textbf{Generalisation to other forms of memory} Diverse dynamical behaviour is observed for different forms of the memory kernel $\gamma$, as represented by the kymograph of the dynamics in one dimension and the respective limit cycle constructed by following the dynamics of the density at a fixed position. (a) Coexisting travelling waves when the system receives feedback from two discrete times in the past with two different strengths (the inset presents the same plot viewed from an different perspective) illustrating a limit cycle that exists in three dimensions. (c) Aperiodic, likely to be spatiotemporally chaotic, dynamics coexisting with more regular oscillations when non-linear terms are incorporated in the active part of the chemical potential.}   
   \label{fig:two-delays}
\end{figure*}

\noindent
For the first example we choose $\gamma$ as
\begin{align}
    \gamma(t) = \alpha_1 \delta(t-\tau_{d,1}) + \alpha_2 \delta(t-\tau_{d, 2}). \label{eq:TwoTimeFeedback}
\end{align}
 When information from a single instant in the past is fed back into the dynamics at the current time, the system evolves to a state where at each point it is possible to construct a Lissajous figure with $\phi(t)$ and $\phi(\br, t-\tau_d)$ at a constant $\br$. When generalised to a two-time delay model, the limit cycles evolves to a closed path in a three dimensional manifold with the axes now being given by the field at times $t$, $t-\tau_{d, 1}$ and $\tau_{d, 2}$, as shown in \cref{fig:two-delays} (b). The choice leads to coexistence of travelling waves with two different wave-numbers and frequencies, as visible in the steady state dynamics in one dimension, see \cref{fig:two-delays} (a). The limit cycles showing phase lag between the current and previous configurations now depends on the position as well. 

In the second example the active contribution to the chemical is
\begin{align}
    \mu_{\rm ac} = \alpha_1 \phi(\br, t-\tau_{d, 1}) + \alpha_2 \left[ \phi(\br, t-\tau_{d, 2}) \right]^2. \label{eq:TwoTimeFeedbackNL}
\end{align}
We find that non-repeating patterns that are likely to be spatio-temporally chaotic arise in the steady state and coexist with more orderly patterns, an observation that is also reflected in the chaotic trajectories that appear instead of a well-defined limit cycle.

For $\tau_{d, 1} < \tau_{d, 2}$, and $\alpha_1>\alpha_2$, the emergent limit cycle at all points in space are identical. For $\alpha_1 < \alpha_2$ however, the behaviours changes qualitatively. In the linear case, see \cref{eq:TwoTimeFeedback}, the limit cycle is deformed close to the defect core (see S.M. for details), while in the non-linear example in \cref{eq:TwoTimeFeedbackNL}, the dynamics is chaotic.

\section{Microscopic motivation} \label{sec:Microscopics}
\noindent
The dynamics in \cref{eq:continuity} and \cref{eq:activeJ} obey number conservation and is well-defined phenomenologically. We show in this section, through an explicit example, that the dynamics can be justified on microscopic grounds.

Consider a collection of particles where the position of the $i$th particle is $\br_i$. The particles are point-like with pair-wise interactions that are derivable from a potential $V(|\br_i - \br_j|)$. In addition to the conservative forces, the particle has a brain, whether chemical or mechanical, that keeps track of its evolution in time. The positions evolves as
\begin{align}
& \frac{\mbox{d} \br_i}{\mbox{d} t} = \frac{1}{\Gamma} \sum_j\bm{F}(\br_i - \br_j) \nonumber \\
& + \frac{\alpha}{\Gamma}\sum_j\int_0^t \mbox{d} t' \gamma(t-t') \bar{\bm{F}}(\br_i(t') - \br_j(t')) + \bm{\zeta}_i, \label{eq:SingleParticleDyn}
\end{align}

where $\bm F$ and $\bar{\bm{F}}$ are conservative forces derivable from interaction potentials $V$ and $\bar{V}$ respectively. The system would be in equilibrium if two conditions are met simultaneously-  $\bar{\bm{F}} = \bm{F}$, and if correlations of the noise $\bm{\zeta}$ are related to $\gamma$ such that the effective temperature is frequency independent. \cref{eq:SingleParticleDyn} represents a system out of equilibrium as the response is colored while the noise is delta-correlated. The dynamics is constructed opposite to a active Brownian particle where the noise has finite correlation associated with a fast relaxing orientational degree of freedom~\cite{Romanczuk2012}. The reverse situation also arises in a colloidal particle in an electric field where fluctuations of counter-ions lead to a coloured noise and a effective temperature that violate fluctuation-dissipation theorem~\cite{Saha2014}. 

We now carry out a systematic coarse-graining procedure with standard assumptions~\cite{DavidDean_1996} to obtain an equation for the equation of motion of the density $\rho = \sum_{i} \delta(\br - \br_i)$ only, see Methods~\cref{sec:MetMicroscopic} for a full derivation of the dynamics. Retaining terms until two orders in gradients we have the following
\begin{align}
    \partial_t \rho(x,t) &=\frac{1}{\Gamma}\partial_x\left[ \rho(x,t)\int \mbox{d} y\rho(y, t) \partial_x V(x - y)\right]  \notag \\
    &+\frac{\alpha}{\Gamma}\partial_x\left[ {\rho(x,t-\tau_d)}   \int \mathrm{d}y \, \rho(y,t-\tau_d) \, \partial_x \bar{V}(x-y) \right]\notag \\
    &+\partial_x[\zeta(x, t) ~\rho^{\frac{1}{2}} (x,t)] + T \partial_x^2 \rho(x,t). 
\end{align}
We have retained terms until quadratic order in the activity $\alpha$, see Methods for the full expression. The higher order terms represent contributions involving gradients of $V$ and $\bar{V}$ with three of more gradients and also cross terms involving both $V$ and $\bar{V}$. The introduction of memory modifies the stochastic terms introducing a coloured noise related to $\gamma$ at order $\partial_x^3$ in the dynamics. The higher order terms represent interesting modifications which can be explored numerically in the future. 


\section{Conclusions} \label{sec:conclusions}
\noindent
The aim of this work is to establish a theoretical framework to explore the basic principles of self-organisation in a collection of particles whose dynamics is controlled by an internal feedback mechanism. To construct a non-Markovian system sustained far from equilibrium, we consider a single conserved scalar field that evolves in two free-energy landscapes. The two contributions to the chemical potential are not trivially additive, as they differ due to the frequency dependence of the mobilities, which take the form of memory kernels in time. This indicates that feedback to the current state arises from a weighted average of past configurations of the density field. The system cannot optimise either of the two free energies and instead evolves to a steady state with persistent currents, which is a clear signature of broken detailed balance.

To make a connection to possible and existing experiments~\cite{holubec2025delayedactiveswimmervelocity,ChicosReview,Kopp_2023_EPL}, we introduce the idea of an active particle with memory, where the active part of its velocity depends on its past history and breaks FDT at the level of a single particle. Note that this manner of constructing an active system bears resemblance to other active matter systems, such as active Brownian particles~\cite{cates2025active} or others~\cite{Saha2014,ShiMahault2026}. The important difference here is that the response, rather than the fluctuations, is coloured in this system - a feature that leads to completely different phenomenology at large length scales such as arrested phase separation and motile patterns. We consider a collection of such particles interacting via physical forces that can be derived from a pairwise interaction potential, in addition to an active contribution to the velocity that is a weighted, history-dependent average of a second set of pairwise forces. An explicit Dean–Kawasaki construction shows that the minimal continuum model motivated above corresponds to a juxtaposition of two chemical potentials, one of which has a frequency-dependent mobility, such that the resulting dynamics is non-variational, as illustrated in \cref{fig:MemorySchematic}(b-e). 

We analyse a simple version of the general model in detail—the Cahn–Hilliard equation with delay—where the feedback from a specific instant in the past of the number density is incorporated into the chemical potential. A numerical exploration reveals dynamical behaviour without any passive counterpart. For strong enough feedback, phase separation dynamics is arrested as the length-scale associated with the domain reaches a parameter-dependent finite value and multiple centres emanating spirals waves are formed in the system. The formation of spirals is associated with a linear instability of the uniform state with complex unstable eigenvalues, a harbinger of oscillatory dynamics and travelling patterns in non-reciprocally interacting soft matter. A mode-truncated approximation to the travelling patterns in one dimension shows that the dispersion relation is actually multi-valued (many possible angular frequencies associated with a single wavelength) although only one of these is pronounced in numerical simulations, opening up the possibility that more complex modes of oscillation is possible with non-linearites or differently constructed initial conditions. 

By generalising the basic model by incorporating feedback from two instants in the past, we demonstrate that the route to the finite-wavenumber instability that leads to the formation of moving patterns in this system is universal in nature: the eigen-modes corresponding to fluctuations of the density and its time derivative collapse at an exceptional point before the real part of the eigenvalues changes sign, thereby destabilising the system via a Hopf bifurcation. Numerical solutions showcase the richness of the model: the limit cycles embedded in three dimensions in \cref{fig:two-delays}(b) show that the model is capable of displaying complexities similar to those reported in non-reciprocal spin systems with more than one species~\cite{weis2025generalized}. The diversity of dynamics seen in the generalization of the model points at the pattern forming abilities of the model beyond travelling waves and spirals.

Some aspects of the model are reminiscent of approaches in the field to introduce activity in a scalar phase separating system - for example by coupling it to two different heat baths~\cite{TwoTempPhysRevE.92.032118,ilker2020phase}. Strikingly, we find that dynamic patterns carrying signatures of activity appear when the field interacts `non-reciprocally' with its earlier configuration making a connection to the non-reciprocal Cahn-Hilliard model~\cite{sahaPRX_2020}. The theory we propose provides a framework for pattern formation involving the minimum number of scalar fields—just one. Furthermore, we note that the single-delay model resembles the dynamics of two coupled non-reciprocal fields and locally traces a limit cycle, while the inclusion of two time scales appears to map onto a multicomponent model~\cite{weis2025generalized}. The idea that temporal feedback in a spatially extended system is a route to novel dynamics away from equilibrium is a concept that can be extended to other forms of active systems such as active solids to create meta-materials. Given the growing recognition of phase separation dynamics in biological and active matter systems~\cite{SAHA20161572,PappuBrangwynne2015,Niebel2019,Franzmann2018,cates2025active}, there are compelling reasons to explore the role of active memory in such systems. However, this work can serve as a foundation for many future studies. In particular, by breaking time-translation invariance, the framework provides a route to examining the coupled effects of activity and ageing in the dynamics of a phase-separating field~\cite{SollichAging}.

An intuitive understanding of how a collection of intelligent agents share information and self-regulate their behaviour has inspired collective control principles in robotic swarms, time-delayed swimmers~\cite{holubec2025delayedactiveswimmervelocity}, and in information-controlled clustering~\cite{ChicosReview}. Our work provides a field-theoretic framework to classify the universal features of such systems. The general principles revealed by our work are likely relevant to self-organisation in biological systems~\cite{Bruckner_Thacik_PNAS}; for example in stress regulation by mechanical feedback~\cite{Gustafson2022}. Finally, future studies of the role of initial conditions will provide an understanding of feedback process and control in active matter~ \cite{FodorControl_PhysRevX.14.011012}.
\\\\
\section{Acknowledgements}
We acknowledge insightful discussions with Ramin Golestanian, Sriram Ramaswamy, Navdeep Rana, and Martin Johnsrud. We are grateful for valuable support and funding by the Max Planck Society. We acknowledge support from the Department of Living Matter Physics.
\\\\
\textbf{Author contributions statement.} V.G. and S.S. designed the research, conducted the research, analyzed the data, and wrote the paper. 
\\\\
\textbf{Data and code availability.} The data, code supporting the main findings of this study are available at \nolinkurl{https://github.com/Vaishnavi-gajendragad/CH-time-delay.git} and its Supplementary Information. Any additional data can be made available upon request.
\\\\
\textbf{Declaration of interests.} The authors declare no competing interests.
\clearpage
\renewcommand{\thefigure}{M\arabic{figure}}
\setcounter{figure}{0}
\begin{widetext}
\section{Methods}
\subsection{Microscopic motivation}\label{sec:MetMicroscopic}
\noindent The microscopic dynamics in 1D is written as:
\begin{align}
    \frac{\mbox{d}X_i}{\mbox{d}t} &= \zeta_i + \frac{1}{\Gamma}\sum_{j = 1}^{N} F(X_i(t) - X_j(t)) + \frac{\alpha}{\Gamma}\sum_{j = 1}^{N} \bar{F}(X_i(t - \tau_d) - X_j(t -\tau_d))
\end{align}
We move forward by considering the evolution of of the following density functional, as done in \cite{DavidDean_1996}, 
\begin{align}
    \rho_i (x, t) = \delta(x - X_i(t) )
\end{align}
and using the definition of the density functional, we can define a arbitrary function $f$ as the following: 
\begin{align}
    f(X_i(t)) =  \int \mbox{d}x~ \rho_i(x, t) f(x) 
\end{align}
Expanding the equation using Ito calculus, we can write down the time evolution as:
\begin{align}
    \frac{\mbox{d}f(X_i(t)}{\mbox{d} t} &= \int dx \rho_i(x, t) \bigg[\partial_x f(x) \frac{1}{\Gamma}\bigg(\sum_j F(x(t) - X_j) + \alpha \sum_j \bar{F}\left(x(t -\tau_d) - X_j(t -\tau_d)\right)  \bigg) \bigg]  \notag \\
    &+\int \mbox{d}x \rho_i(x, t) \bigg[\partial_x f(x) \zeta_i(t) \bigg] + T\int \mbox{d} x \rho_i(x, t) \bigg[ \partial_x^2 f(x) \bigg] \label{dfdt}
\end{align}
Here we can isolate the $\rho_i(x, t)$, using integration by parts in Eq. \eqref{partiat_rho1}:

\begin{align}
    \partial_t \rho_i(x,t) &= -\partial_x\left[ \frac{1}{\Gamma}\rho_i (x,t) \sum_j  F(x - X_j (t))  +  \frac{\alpha}{\Gamma} \rho_i (x,t)  \sum_j  \bar{F} [x(t - \tau_d) - X_j(t - \tau_d)]  \right] \notag \\
    &-\partial_x[\zeta_i \rho_i (x,t)] + T \partial_x^2 \rho_i(x,t) \label{partiat_rho1}
\end{align}
We first consider the terms proportional to $\alpha$ in the above equation and expand it as the following:
\begin{align}
    \rho_i(x, t) &=  \delta(x - X_i(t - \tau_d) - \Delta X_i(t, \tau_d)) \\
    \Delta X_i(t, \tau_d) &= X_i(t) - X_i(t - \tau_d)
\end{align}
and we expand $\rho_i(x, t)$ as the following using Ito calculus:
\begin{align}
    \rho_i(x, t) = \bigg[1 - \Delta X_i(t, \tau_d) \partial_x + \frac{\langle \Delta  X_i(t, \tau_d) ^2 \rangle}{2} \partial_x^2   \bigg] \rho_i(x, t - \tau_d) \label{rho_rhotd}
\end{align}
Here we write down dynamics of $\Delta X_i$ explicitly 
\begin{align}
    \Delta X_i(t, \tau_d)  = \frac{1}{\Gamma }\int_{t - \tau_d}^{t} \mbox{d}t'\bigg(  \sum_i^N F(X_i(t') - X_j(t')) +  \alpha \sum_i^N \bar{F}(X_i\left(t' - \tau_d) - X_j(t' - \tau_d)\right)\bigg) + \int_{t - \tau_d}^{t}\mbox{d}t' \zeta_i(t') \label{DeltaX}
\end{align} 
and expand it linear order of $\tau_d$ and $\alpha$. Where, $\bar{\zeta}_i= \int_{t - \tau_d}^{t}\mbox{d}t' \zeta_i(t')$. 
\begin{align}
    \Delta X_i(t, \tau_d) \approx \frac{\tau_d}{\Gamma}  \sum_j F(X_i(t) - X_j(t) ) + \alpha \frac{\tau_d}{\Gamma}  \sum_j \bar{F}(X_i(t - \tau_d) - X_j(t - \tau_d)  ) + \bar{\zeta}_i(t) \label{DeltaX_app}
\end{align}
Now consider the term proportional to $\alpha$ in \eqref{partiat_rho1}, substituting the expression of $\rho_i(x, t)$ Eq. \eqref{rho_rhotd} we obtained above, and expressing it in terms of potential $\sum_j \bar{V}[x(t - \tau_d) - X_j(t - \tau_d)]$, we get:
\begin{align}
    \rho_i (x,t)  \sum_j  \bar{F} [x(t - \tau_d) - X_j(t - \tau_d)] & = - \rho_i(x,t-\tau_d) \sum_j \int \mathrm{d}y \, \rho_j(y,t-\tau_d) \, \partial_x \bar{V}(x-y) \notag \\
    & \qquad - \Delta X_i(t,\tau_d) \, \rho_i(x,t-\tau_d) \sum_j \int \mathrm{d}y \, \rho_j(y,t-\tau_d) \, \partial_x^2 \bar{V}(x-y) \notag \\
    & \qquad + T \tau_d \, \rho_i(x,t-\tau_d) \sum_j \int \mathrm{d}y \, \rho_j(y,t-\tau_d) \, \partial_x^3 \bar{V}(x-y)
\end{align}
In the above we substitute the expression for $\Delta X_i(t, \tau_d)$, and bring it all together as the following:
\begin{align}
    \partial_t \rho_i(x,t) &= -\partial_x\left[ -\frac{1}{\Gamma}\rho_i (x,t)\sum_j \int \mbox{d} y\rho_j(y, t) \partial_x V(x - y)\right]  \notag \\
    &-\partial_x\left[-  \frac{\alpha}{\Gamma}  \rho_i(x,t-\tau_d) \sum_j \int \mathrm{d}y \, \rho_j(y,t-\tau_d) \, \partial_x \bar{V}(x-y) \right]\notag \\
    &-\partial_x[\zeta_i \rho_i (x,t)] + T \partial_x^2 \rho_i(x,t) \notag\\
    &- \partial_x\left[- \frac{\alpha T\tau_d}{\Gamma}  \rho_i(x, t- \tau_d)   \sum_j  \int \mathrm{d}y \, \rho_j(y,t-\tau_d) \, \partial_x^3 \bar{V}(x-y)\right] \notag\\
    &-\partial_x\left[ \frac{- \alpha \tau_d}{\Gamma}\sum_j \int \mathrm{d}y \, \rho_i(x,t-\tau_d) \rho_j(y,t-\tau_d) \partial^2_x\bar{V}(x-y) \bigg[  \sum_j \int \mbox{d} z\rho_j(z, t) \partial_x V(x - z) \bigg] \right] \notag \\
    &-\partial_x\left[\alpha^2 \frac{\tau_d}{\Gamma}\sum_j \int \mathrm{d}y \,  \rho_i(x,t-\tau_d)   \rho_j(y,t-\tau_d) \partial^2_x \bar{V}(x-y) \bigg[   \sum_j \int \mbox{d} k\rho_j(k, t - \tau_d) \partial_x \bar{V}(x - k)\bigg] \right] \notag \\
    &- \partial_x\left[ \sum_j \int \mathrm{d}y \,  \rho_i(x,t-\tau_d)   \rho_j(y,t-\tau_d) \partial^2_x \bar{V}(x-y) \bar{\zeta}(t) \right] 
\end{align}
\textcolor{black}{It is important to note here that we expand $\rho_i(x, t)$ instead of $F'$ to avoid non-local temporal couplings which cannot be reduced to local time derivative of $\rho$ unless additional approximations are made.} Restricting it to second order gradients, summing over $i$, using the definition $\rho(x, t) = \sum_i \delta(x - X_i(t) )$.
Additionally, the noise term in the above, which is coupled with $\rho(x, t)$, is resolved as $\partial_x ( \zeta(x, t) \rho^{1/2}(x, t))$, following the same procedure outlined in \cite{DavidDean_1996}.
\begin{align}
    \partial_t \rho(x,t) &=\frac{1}{\Gamma}\partial_x\left[ \rho(x,t)\int \mbox{d} y\rho(y, t) \partial_x V(x - y)\right]  \notag \\
    &+\frac{\alpha}{\Gamma}\partial_x\left[ {\rho(x,t-\tau_d)}   \int \mathrm{d}y \, \rho(y,t-\tau_d) \, \partial_x \bar{V}(x-y) \right]\notag \\
    &+\partial_x[\zeta(x, t) ~\rho^{\frac{1}{2}} (x,t)] + T \partial_x^2 \rho(x,t) 
\end{align}
Rewriting the equation in terms of free energy functional as given by  \cite{DavidDean_1996}, we arrive at the final coarse grained equation
\begin{align}
    \partial_t \rho(x,t) &=\frac{1}{\Gamma}\partial_x\left[ \rho(x,t)\partial_x\bigg(\frac{\delta  \mathcal{F}}{\delta \rho} \bigg)\right]  \notag \\
    &+\frac{\alpha}{\Gamma}\partial_x\left[ {\rho(x,t-\tau_d)} \partial_x\bigg( \frac{\delta \mathcal{F}'}{\delta \rho}\bigg) \bigg|_{t - \tau_d}\right]\notag \\
    &+\partial_x[\zeta(x, t) ~\rho^{\frac{1}{2}} (x,t)] + T \partial_x^2 \rho(x,t) 
\end{align}
\end{widetext}
\subsection{Fluctuation-Dissipation ratio}\label{sec:MetFDT}
\noindent The Fourier transform of the correlation function is given by
\begin{align}
    C(\bm{q}, \bm{q}',\omega,\omega') =\langle \phi(\boldsymbol{q}, \omega) \phi(\boldsymbol{q}', \omega') \rangle \notag \\ =C_{\rm S}(\boldsymbol{q}, \omega)  \delta^d (\boldsymbol{q} + \boldsymbol{q}') \delta (\omega + \omega')
\end{align}
Substituting the expression for $\phi(\boldsymbol{q}, \omega)$ from the linearized equation \eqref{lin_phiqom} the correlation function becomes
\begin{align}
     C(\bm{q}, \bm{q}',\omega,\omega')  = \frac{2q^2 D }{|i\omega + q^2(D_{eff}  + \alpha \gamma(\omega))|^2}
\end{align}
Knowing $\gamma(\omega) = \exp(i\omega \tau_d)$ we can split it into its real and imaginary components $\gamma = \gamma'(\omega) + i \gamma''(\omega)$ to simplify the above expression to obtain
\begin{align}
    C_{\rm S}(\boldsymbol{q}, \omega) = \frac{2D q^2}{(\omega + \alpha q^2\gamma'')^2 + {q}^4(D_{eff} + \alpha \gamma')^2}
\end{align}
The response function is calculated the usual way by perturbing the free energy by $- h \phi$,
\begin{align}
    \phi(\boldsymbol{q}, \omega) &= \frac{q^2 h + i \boldsymbol{q} \cdot \boldsymbol{\zeta} }{i(\omega + \alpha q^2 \gamma'') +  q^2(D_{eff}  + \alpha \gamma')}  \\
    R(\boldsymbol{q}, \omega) &= \frac{q^2 (\omega + \alpha q^2 \gamma'') }{(\omega + \alpha q^2 \gamma'')^2 +  {q}^4(D_{eff}  + \alpha \gamma')^2}
\end{align}
Hence the fluctuation dissipation ratio becomes
\begin{align}
    \frac{\omega C}{2D R} = \frac{\omega}{\omega + \alpha q^2 \gamma''}
\end{align}
\subsection{Linear stability analysis}\label{sec:MetLinStab}
\begin{figure}[h!]
    \centering
    \includegraphics[width=0.9\linewidth]{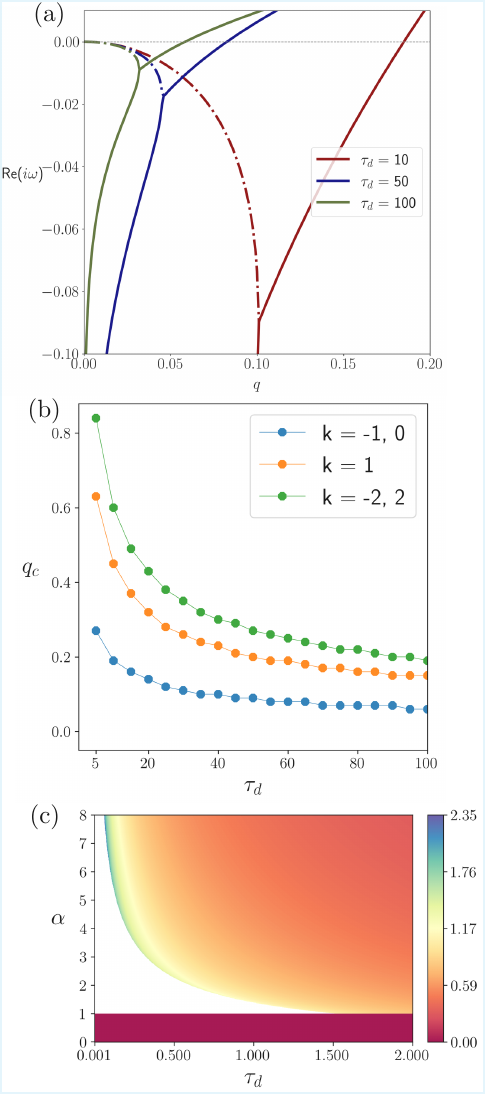}
    \caption{\textbf{Critical wavenumber associated with instability}: 
    Panel (a) illustrates finite wavenumber instability $q_c$ at different values of $\tau_d$.
    Panel (b) illustrates $q_c$ versus $\tau_d$ at $\alpha = 4.0$.
    Panel (c) shows $q_c$ as a function of $\alpha$ and $\tau_d$. White regions correspond to parameter values where no instability occurs. For $\alpha < 1$, passive phase separation sets in with instability at $q_c = 0$. As $\tau_d \to 0$, $q_c$ increases, whereas for larger $\tau_d$, it remains finite but shifts toward zero. }
    \label{Mfig:qc}
\end{figure}
\begin{figure*}[h]
    \centering
    \includegraphics[width=0.55\linewidth]{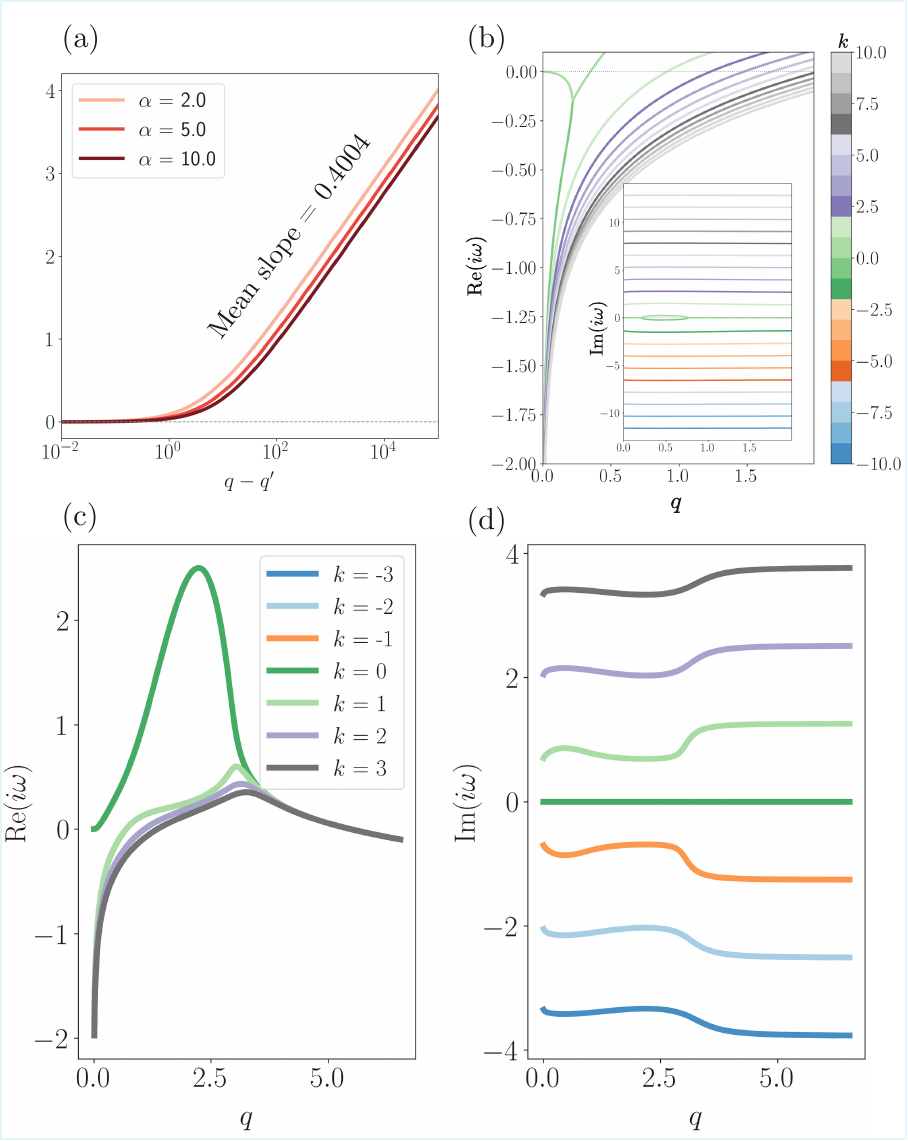}
    \caption{(a) Semi-log plot illustrating the decay of branches ($|\text{Re}(i\omega)|$) for $\tau_d = 5.0, \kappa = 1$, where $q'$ is point at which the branches becomes stable again ($\text{Re}(i\omega) < 0$) (b) Plot (with the inset showing the corresponding imaginary components) illustrates the emergence of multiple branches (for $k \in (-10, 10)$), highlighting the multi-valued nature of the characteristic equation.  (c), (d) Linear stability analysis for $\alpha < 0$, illustrating the absence of exceptional points, Hopf bifurcation for $k = 0, -1$}
    \label{Mfig:Branches}
\end{figure*}
\noindent The transcendental equation obtained by perturbing the homogenous state is solved using Lambert W function. Where $W_k$ satisfies the equation  $W_k(z) e^{W_k(z)} = z$, for a complex number $z = x + iy$. The roots $i \omega$ can be written as
\begin{align}
    i\omega = \frac{W_k \bigg( -\alpha \tau_d {q}^2 e^{\tau_d(  \kappa {q}^4 - {q}^2)} \bigg)}{\tau_d} +  ({q}^2  -\kappa {q}^4) \label{lambda_W1},
\end{align}

showing a natural split into two parts, where the first contribution stems from activity while the second contribution is identical to passive model B. The roots are labelled by the branches $k$ of the $W$ functions following usual conventions. An essential property of this function is that the real values  of the function lie in $-1 \leq \text{Re}(W(x)) < 0$ for $x \in [-1/e , 0) $, $W(x=0) = 0$ and $\text{Re}(W(x)) > 0$ for $x > 0$. However, the first term of our equation lies outside this domain $(< -1/e)$, making it inaccessible through to the function, and are evaluated numerically. 

\subsection{Numerical scheme}\label{sec:MetNumerics}
\noindent We use a pseudo-spectral algorithm to integrate the model numerically in both one-dimensional (box size $L$, resolution $N$) and two-dimensional square periodic box (box size $L_x = L_y = L, N_x = N_y = N$). 
For all simulations shown in \cref{fig:DynStates}, we use $L = 50$ and $N = 512$, and integrate the system for long times ($\sim 5 \times 10^6$ to $ 1 \times 10^7$ time steps). The time step is $\mathrm{d}t = 10^{-5}$ for $\alpha = 2$ and for $\alpha = 4$ with $\tau_d \in (0.01, 16.66)$, except at $\alpha = 4$, $\tau_d = 16.66$, where a smaller time step $\mathrm{d}t = 10^{-6}$ is used. For $\alpha = 8$, we use $\mathrm{d}t = 10^{-6}$ for all values of $\tau_d$.
 
For time marching we use first-order exponential time differencing scheme \cite{COX2002430}. For simulations with noise we use first-order stochastic exponential time differencing scheme \cite{Johnsrud2025Nov}.

\clearpage
\bibliography{biblio}
\newpage
\section{Supplementary}
\noindent This supplementary document contains details of analytical calculations and numerical procedures connected to the results presented in the main file. \ref{video_captions} contains details and descriptions of supplementary videos of pattern formation in the system, \ref{LS_reduced} consists of details of linear stability analysis in the reduced description and the entropy production at exceptional points. \ref{OMT} contains the derivation of dispersion relation and entropy using one-mode truncation. \ref{qselect} contains the numerical details of wavenumber selection, with results showing examples of low and large variance. In \ref{2td}, we present different cases of the model with multiple time delays and the results of linear stability analysis around the homogenous state. 
\subsection{Captions of Videos}\label{video_captions}
\begin{itemize}
    \item \textbf{Movie S1, Travelling wave with defects}: For $\alpha = 4.0, \tau_d = 2.083$, we initially observe the formation of a spiral, whose radius unidirectionally (upper right) grows as big as the box size, leading to its eventual destruction due to periodic boundary conditions. The system resolves into travelling waves with persistent defects in steady state. 
    \item \textbf{Move S2, Travelling waves}: For $\alpha = 8.0, \tau_d = 0.07$, we observe initial transition to formation of travelling wave with defects, but the defects get ``untangled" and resolve into simpler travelling waves. 
    \item \textbf{Movie S3, Spiral}: For $\alpha = 4.0, \tau_d = 4.17$, We observe the formation spiral in the bottom right corner of the box, which grows radially outwards in all directions until it spans the entire box and hence persisting in the steady state.  We observe that this is one of the key factors in the persistence of spirals in steady state, along with formation of multiple spirals that stabilize each other, as presented in in Fig. 3.
\end{itemize}

\subsection{Linear stability analysis in reduced description}\label{LS_reduced}
\noindent
The field at an earlier time can be expanded as
\begin{align} \label{eq:FieldApprox}
    \phi(\boldsymbol{r}, t - \tau_d) \approx \phi(\boldsymbol{r}, t) - \tau_d \frac{\partial  \phi(\boldsymbol{r}, t)}{\partial t} + \frac{\tau_d^2}{\textcolor{black}{2}} \frac{\partial^2  \phi(\boldsymbol{r}, t)}{\partial t^2}
\end{align}
Substituting the expression in the Cahn-Hilliard equation in the main paper, see Eq. 12 of the main text, we obtain the following equation for perturbations in Fourier space
\begin{align}
    &(1 - \alpha \tau_d q^2)\partial_t \delta \phi  + \frac{\alpha \tau_d^2 q^2}{2} \partial_t ^2 \delta \phi \nonumber \\
    & + q^2( (a + \alpha)  + \kappa q^2 )\delta \phi  = \sqrt{2D} i \bm{q} \cdot {\zeta}  \label{2nd_order_exp}
\end{align}
As presented in the main we identify the coefficients of the above second-order langevin equation, as the following:
\begin{align}
    & \Gamma'(q) =  (1 - \alpha \tau_d q^2), \nonumber \\
& m(q) = \frac12 \alpha \tau_d^2 q^2, \nonumber \\
& K(q) = q^2( (a + \alpha)  + \kappa q^2 ). \label{eq:ReducedEq}
\end{align}
In Fourier space, this correlation is given by:
\begin{align}
    \langle \zeta(\boldsymbol{q}, \omega) \zeta(\boldsymbol{q}', \omega')\rangle = (2\pi)^{d + 1} \delta (\omega + \omega') \delta (\boldsymbol{q} + \boldsymbol{q}')
\end{align}
We identify the mass, damping, and spring constant as discussed in Section V A of the main text. Introducing the dynamical matrix $\mathbb{M}$ as following
\begin{align}
    \begin{pmatrix}
        \partial_t \delta \phi \\
        \partial_t v
    \end{pmatrix} 
    = \mathbb{M} \cdot \begin{pmatrix}
        \delta \phi \\
        v
        \end{pmatrix}, 
\end{align}
where 
\begin{align}
    \mathbb{M} = \begin{pmatrix}
        0 & 1 \\
        \frac{-2q^2(a + \alpha  + \kappa q^2) }{\alpha q^2 \tau_d^2} & \frac{-2(1 - \alpha \tau_d q^2)}{\alpha q^2 \tau_d^2}
    \end{pmatrix}\begin{pmatrix}
        \delta \phi \\
        v
        \end{pmatrix},
\end{align}
yielding the eigenvalue equation
\begin{align}
    \lambda^2 + \frac{2(1 - \alpha \tau_d q^2)}{\alpha \tau_d^2 q^2} \lambda + \frac{2((a + \alpha)  + \kappa q^2)}{\alpha \tau_d^2} = 0 
\end{align}
with the eigenmodes 
\begin{align} 
    \lambda_{\pm} &= -\frac{(1 - \alpha \tau_d q^2)}{\alpha \tau_d^2 q^2} \pm \frac{1}{2}{\sqrt{\Delta}}, \label{lm_red}\\
    \Delta &= \frac{4(1 - \alpha \tau_d q^2)^2}{\alpha^2 \tau_d^4 q^4} - 8\frac{a + \alpha + \kappa q^2}{\alpha \tau_d^2} \notag  \label{eq:eigenLambdaPm}
\end{align}
To understand the $q \longrightarrow 0$ limit of the above roots, we can expand expressions for small $q$. We get:
\begin{align}
    \lambda_{+}(q) &= -(a+\alpha)\,q^2 + \mathcal{O}(q^4), \\
\lambda_{-}(q) &= -\frac{2}{\alpha\tau_d^2 q^2} + \frac{2}{\tau_d} + (a + \alpha) q^2 + \mathcal{O}(q^4).
\end{align}
Hence at $q \longrightarrow 0$:
\begin{align}
    \lim_{q \longrightarrow0} \lambda_{+} = 0 \\
    \lim_{q \longrightarrow0} \lambda_{-} = -\infty
\end{align}
\noindent The properties of $\lambda_+, \lambda_-$ at $q \rightarrow 0$ mimics the behaviour of the branches $k = 0, -1$ from the transcendental equation. Additionally, from the figure we see the occurrence of the first Hopf bifurcation point. 

The right eigenvectors are defined as $\mathbb{M} \cdot \boldsymbol{v}_{\pm} = \lambda \boldsymbol{v}_{\pm}$.
    \newline 
    Where $\boldsymbol{v_+}, \boldsymbol{v_-}$ are eigenvectors of the eigenvalues $\lambda_+, \lambda_-$ respectfully. 
We can  write the eigenvectors as
\begin{align}
    \boldsymbol{v_{\pm}} &= \frac{1}{N_{\pm}} \begin{pmatrix}
        \Lambda_{\pm} \\
        2
    \end{pmatrix} \\
    \Lambda_{\pm} &= \alpha \tau_d^2( -1 + q^2 \alpha \tau_d) \pm\sqrt{\Delta}
\end{align}
And the normalised eigenvectors are:
\begin{align}
    N_\pm = q^2 \alpha (a + \alpha + q^2 \kappa) \tau_d^2 
\sqrt{ 4 + \left\lvert \frac{\Lambda_\pm}{q^2 \alpha (a + \alpha + q^2 \kappa) \tau_d^2} \right\rvert^2 },
\end{align}
Similarly, the left eigenvectors are calculated as $\boldsymbol{v}_L \cdot \mathbb{M} = \lambda\boldsymbol{v}_L$.
The phase difference between the two Eigenvectors as the following:
\begin{align}
    \Theta = \cos^{-1}(\boldsymbol{v}_+ \cdot \boldsymbol{v}_-)  
\end{align}
From the figure, we clearly see that after the exceptional point, $\mbox{Re} (\lambda_+) = \mbox{Re} (\lambda_-)$, but $\mbox{Im}(\lambda_+) \neq \mbox{Im}(\lambda_+)$. In fact, the two eigenvalues are complex conjugates of each other. Hence the eigenvectors have a phase difference.
\subsection{Reduced description for a general memory kernel}
\noindent Consider the following reduced, linearised description with a general memory kernel $\gamma$:
\begin{align}
    i\omega = -q^2(a + \kappa q^2) - \alpha q^2\gamma(i\omega) 
\end{align}
Expanding $\gamma(i\omega)$ using Taylor series:
\begin{align}
    \gamma(i\omega) = \gamma(0) + i\omega \gamma'(0) + \frac{1}{2}(i\omega)^2 \gamma''(0)
\end{align}
And substituting this in the original expression, we get the following quadratic equation:
\begin{align}
    i\omega\Gamma'(q) + m(q)(i\omega)^2 &= K(q)
\end{align}
Where 
\begin{align}
    \Gamma'(q) &= 1 - \alpha q^2 \gamma'(0) \\
    m(q) &= \frac{\alpha q^2 \gamma''(0)}{2} \\
    K(q) &= -q^2(a + \kappa q^2 +\alpha \gamma(0))
\end{align}
When the memory kernel is of the form of a single time delay $\gamma(\omega) = \exp(i\omega \tau_d)$, the constants have the form given in Eq.~\eqref{eq:ReducedEq}. When there are multiple memory kernels of the form $\gamma(\omega) = \exp(i\omega \tau_{d_1}) + \exp(i\omega \tau_{d_2}) $, the constants have the following form:
\begin{align}
    \Gamma'(q) &= 1 - \alpha q^2(\tau_{d_1}  + \tau_{d_2}) \notag \\
    m(q) &= \frac{\alpha q^2 (\tau_{d_1}^2  + \tau_{d_2}^2)}{2} \notag \\
    K(q) &= -q^2(a + \kappa q^2 +\alpha )
\end{align}
\subsection{Entropy production at the exceptional point}
\noindent 
The entropy production rate in the system is
\begin{align}
    S = -\frac{1}{D} \int d \mathbf{r} \langle \mu_A \dot{\phi} \rangle \label{entropy_prod}
\end{align}
We consider linearised perturbations around homogenous state $(\bar{\phi} + \delta \phi)$ and Taylor expand the active potential as given in the previous section. We find that the rate of entropy production for fluctuations around a uniform state is
\begin{align}
    \label{eq:EntUniform}
     S &= -\frac{1}{D} \int d \mathbf{r}\alpha \langle \partial_t \phi(\br, t) \delta \phi(\br, t - \tau_d) \rangle,  \nonumber \\
&= -\frac{\alpha}{D} \int d \mathbf{r}\alpha \langle v (\br, t) \delta \phi (\br, t) \rangle   \nonumber \\
&  + \frac{\alpha \tau_d}{D} \int d \mathbf{r} \langle v(\br, t)^2 \rangle \nonumber \\
& - \frac{\alpha \tau_d^2}{2D} \int d \mathbf{r} \langle \partial_t v(\br, t) v(\br, t) \rangle
\end{align}
Where $v = \partial_t \delta \phi (\br, t)$. The first term on the RHS of Eq.~\ref{eq:EntUniform} vanishes in the steady state, while the second is a positive quantity. 
Taking the Fourier transform in time, 
\begin{align}
    \delta \phi(\boldsymbol{q}, \omega) =  \frac{\sqrt{2Dq^2} \zeta(\boldsymbol{q}, \omega)}{m(i\omega - \lambda_-)(i\omega - \lambda_+)},
\end{align}
where $\lambda_+, \lambda_-$ are the eigenvalues in Eq.~\eqref{eq:eigenLambdaPm}. The first term, which simplifies as:
\begin{align}
    \langle v \delta \phi \rangle = \sqrt{2Dq^2} \int \frac{i\omega}{(i\omega -\lambda_+)(i\omega - \lambda_-)} \mbox{d}\omega
\end{align}
and the second term:
\begin{align}
\langle v^2 \rangle  = \frac{ {2 Dq^2} }{m^2 } (2\pi)^2\int  \frac{\omega^2}{(\omega^2 +\lambda_+^2)(\omega^2 + \lambda_-^2)} \mbox{d}\omega 
\end{align}
The third term vanishes since it is odd ordered in $\omega$.
\newline \noindent At the exceptional point $\lambda_- = \lambda_+ = \lambda$, while the second term simplifies to:
\begin{align}
    \langle v^2 \rangle  &= \frac{ {-2 Dq^2} }{m^2 } (2\pi)^2\int  \frac{\omega^2}{(\omega^2 +\lambda^2)^2} \mbox{d}\omega \notag \\
    &= \frac{ {2Dq^2} }{m^2 } (2\pi)^2 \frac{\pi}{2\lambda}  
\end{align}
The first term diverges:
\begin{align}
    \langle v \delta \phi \rangle &= \sqrt{2Dq^2} \int \frac{i\omega}{(i\omega - \lambda)^2} \mbox{d}\omega \notag \\
    &= \int \frac{1}{i\omega - \lambda} \mbox{d}\omega 
    + \underbrace{\int \frac{\lambda}{(i\omega - \lambda)^2} \mbox{d}\omega}_{\text{Divergent contribution}} 
\end{align}
Hence at the exceptional point, the entropy production diverges. 

\subsection{One mode truncation}\label{OMT}
\noindent 
In a system of size $L$, we truncate the Fourier representation of $\phi$ to the fundamental wavenumber $q_0 = 2\pi/L$ only to write it as 
\begin{align}
    \phi = \bar{\phi} + a(t)\exp^{iq_0 x} + a^*(t) \exp^{-i q_0 x}, \label{eq:onemOde}
\end{align}
\begin{align}
    \dot{a^*} &=& q_0^2 a^* \bigg( 1 - \kappa q_0^2 - 3|a|^2\bigg)- \alpha q_0^2 a^*(t - \tau_d)
\end{align}
Now, substituting $a = A \exp{(-i \Omega t)}, \Longrightarrow |a|^2 = A^2$, we get:
\begin{align}
    -A i \Omega \exp{(-i \Omega t)}  &=& q_0^2 A \exp{(-i \Omega t)} \big( 1 -  \kappa q_0^2 - 3A^2\big) \notag \\
    &-& \alpha q_0^2 A \exp{(-i \Omega (t - \tau_d) )}
\end{align}
Which can be simplified as:
\begin{align}
    -i \Omega &=& q_0^2 \big( 1 -  \kappa q_0^2 - 3A^2\big) \notag \\
    &-& \alpha q_0^2 \big( \cos(\Omega\tau_d) + i \sin(\Omega\tau_d) \big)
\end{align}
Now equating the imaginary parts, we get the following dispersion relation:
\begin{align}
    \Omega &= \alpha q_0^2 \sin(\Omega \tau_d) \label{sin_om}
\end{align}
and the real parts, we get the following amplitude condition:
\begin{align}
     A^2 = \frac{1}{3} \bigg( 1 - \kappa q_0^2 - \alpha \cos(\Omega \tau_d) \bigg)
\end{align}
A interesting scaling property becomes apparent when when we multiply the \eqref{sin_om}, with $\tau_d$
\begin{align}
    \Omega^{\prime}  = \alpha^{\prime} q_0^2 \sin (\Omega^{\prime})
\end{align}
\begin{align}
    A^2 = \frac{1}{3} \bigg( 1 - \kappa q_0^2 - \alpha^{\prime}/ \tau_d \cos(\Omega^{\prime}) \bigg)
\end{align}
Where $\Omega^{\prime} = \Omega \tau_d$ and $ \alpha^{\prime} = \alpha \tau_d $.
\noindent
For the trivial solution $\Omega = 0$, the roots are physical only until a certain point. To calculate this critical $\alpha$, we take $A = 0$, and obtain the following condition:
\begin{align}
    \alpha_c = 1 - \kappa q_0^2 
\end{align}
There this critical $\alpha_c$ for the trivial solution is independent of $\tau_d$ and hence constant for all values of $\tau_d$. 
\newline
\noindent
As clearly seen in the main figure, there are bifurcation points (marked by red circles) at which more than one root $(\Omega)$ is possible. The points at which \eqref{sin_om} becomes multi-valued for a range of $\alpha$ and $\tau_d$ values is where line $\Omega$ is tangent to $f(\Omega) = \alpha q_0^2 \sin(\Omega')$, say $\bar{\Omega}$, which is when the slope are equal to each other. 
The slope of $f(\Omega)$ at $\bar{\Omega}$ 
\begin{align}
    f'(\bar{\Omega}) = \alpha q_0^2 \tau_d \cos(\bar{\Omega} \tau_d) = 1
\end{align}
\begin{align}
    \cos(\bar{\Omega} \tau_d)  = \frac{1}{\alpha q_0^2 \tau_d} \label{slope_cond} \noindent \\
    \bar{\Omega} = \frac{1}{\tau_d}\cos^{-1}\bigg(\frac{1}{\alpha q_0^2 \tau_d} \bigg)
\end{align}
Hence the general solution for when the bifurcations occur:
\begin{align}
    \bar{\Omega}_n = \frac{1}{\tau_d}\cos^{-1}\bigg(\frac{1}{\alpha q_0^2 \tau_d} \bigg) + 2\pi n
\end{align}
\subsection{Entropy production in the travelling wave state}
\noindent
\begin{figure*}
    \centering
    \includegraphics[width=0.8\linewidth]{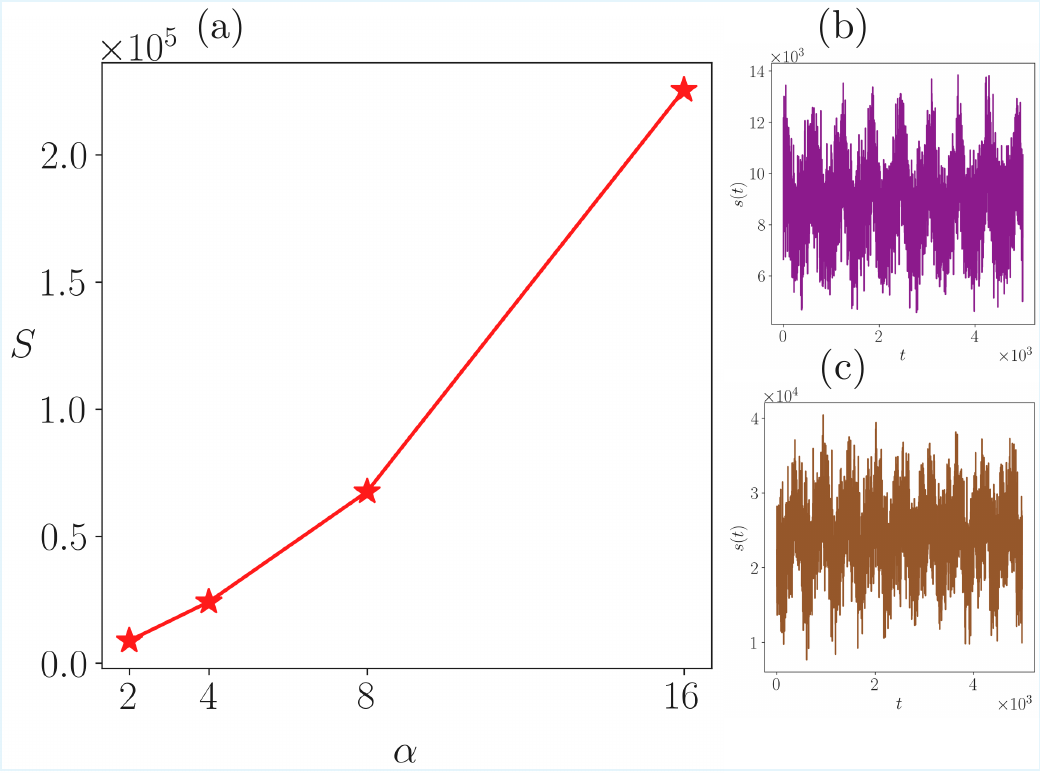}
    \caption{\textbf{Total entropy production as a function of $\alpha$} for $N = 1024, L = 100, D=0.0001, \tau_d = 5.0$. 
    Numerical procedure: For each value of $\alpha$, 5 independent realizations are initiated. $S$ of steady state of each realization is calculated and averaged.  (b) Ensemble averaged Entropy production density $s(t)$ for $\alpha = 2.0$ (c) for $\alpha = 4.0$ as function of $t$}
    \label{Ent_fig}
\end{figure*}
We know, the entropy production is given by: 
\begin{align}
    S = -\frac{1}{D} \int \langle \mu_A \dot{\phi} \rangle d \mathbf{r} \label{entropy_prod}
\end{align}
Using $\mu_A = \alpha \phi(\boldsymbol{r}, t - \tau_d)$, the local entropy production density is given by:
\begin{align}
    s = \lim_{\tau \rightarrow \infty}\frac{-\alpha}{D \tau } \int_0^\tau dt \langle  \phi(\boldsymbol{r}, t - \tau_d)~ \dot{\phi} (\boldsymbol{r}, t)\rangle ~(\boldsymbol{r}) 
\end{align}
Applying the travelling wave solution of the following form:
\begin{align}
    \phi = |A| \sin(q_0 x - \Omega t)
\end{align}
The integrand simplifies as the following, where $\psi = q_0 x - \Omega t$
\begin{align}
    \phi(x, t - \tau_d)\dot{\phi}(x, t) = -\Omega |A|^2 \sin(\psi + \Omega \tau_d)  \cos ( \psi)
\end{align}
and using the product to sum identity to simplify the above and taking the integral, we get:
\begin{align}
    s(t) = \frac{1}{2}\int_0^L \sin(2\psi + \Omega \tau_d) dx + \frac{1}{2}\int_0^L \sin(\Omega \tau_d) dx
\end{align}
The first integral vanishes due to the periodicity of $q_0 = 2\pi n/L $, $n = 0 , 1, 2, \ldots N -1$, $\Longrightarrow q_0L = 2\pi n$.
\begin{align}
    s(t) = \frac{\alpha L |A|^2}{2 D}  \Omega \sin(\Omega \tau_d) 
\end{align}
And the total entropy production is given by:
\begin{align}
    S &= \frac{\alpha L |A|^2}{2D} \Omega \lim_{T \to \infty} \frac{1}{T}\int_0^T \sin(\Omega \tau_d) dt \notag \\ 
    S &= \frac{\alpha L |A|^2}{2D} \Omega \sin(\Omega \tau_d)
\end{align}
\subsection{Wavenumber selection}\label{qselect}
\begin{figure*}[h]
    \centering
    \includegraphics[width=\linewidth]{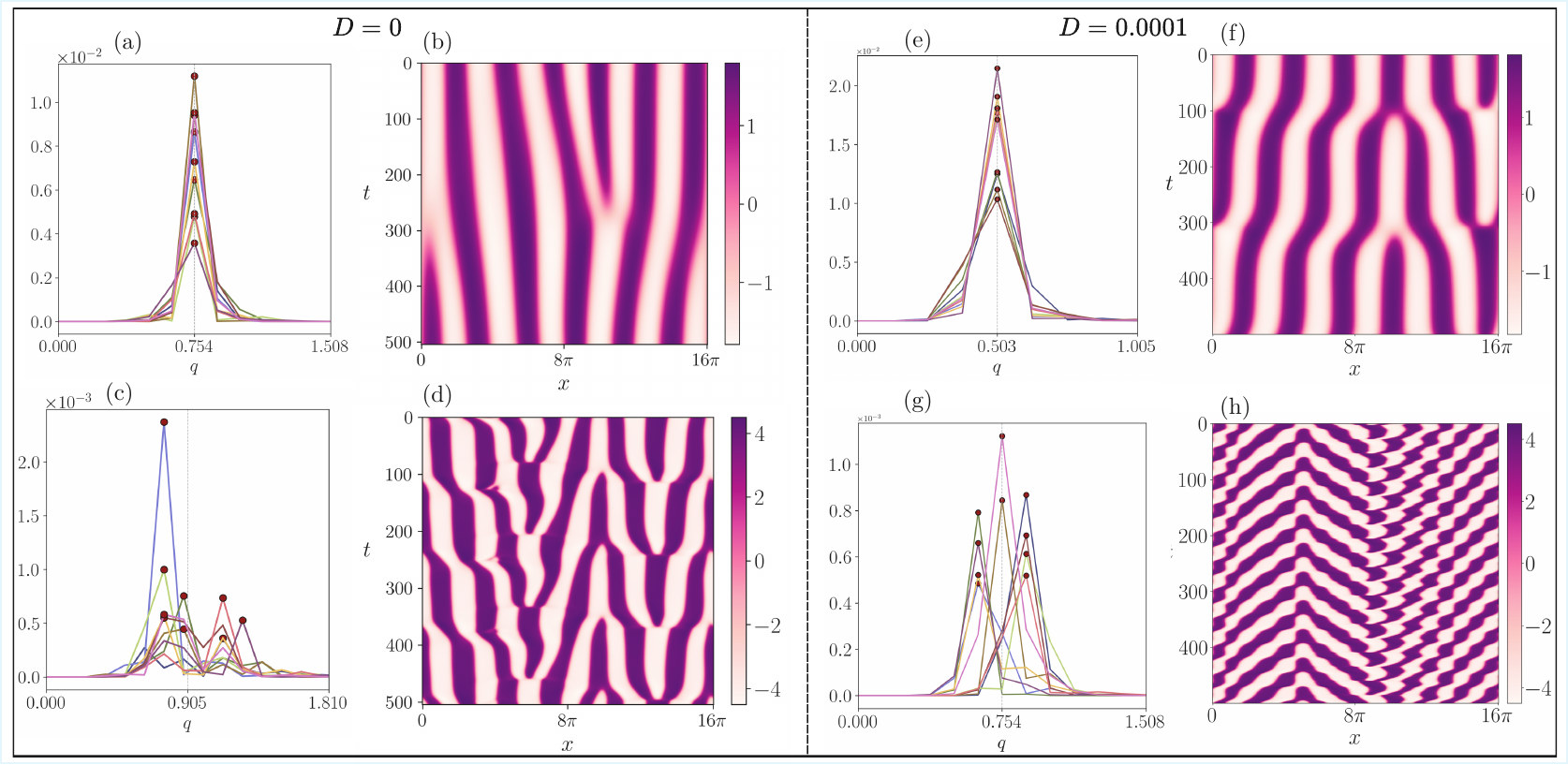}
    \caption{Panels (a) and (c) show static structure factors for $\alpha = 2.0, \tau_d = 10.0$ and $\alpha = 16, \tau_d = 2.5$, respectively, with $N = 512, L = 50$ (corresponding kymographs in (b) and (d)). Panels (e) and (g) show $S_q$ for $\alpha = 2.0, \tau_d = 20.0$ and $\alpha = 16, \tau_d = 2.5$ in the presence of noise ($D = 0.0001$), with kymographs in (f) and (h). Together, these examples span the range of wavenumber selection, from the lowest to highest observed variances.}
    \label{qselect_app}
\end{figure*}

\noindent For every a pair of $\alpha$ and $\tau_d$ there is peak  of structure factor (Fourier transform of two-point correlation function of $\phi$) at a certain wavenumber $q_0$, at steady state which corresponds to the coarsening length scale. 
To ensure the selection of wavenumber is initial condition independent, we run 10 independent realizations for each pair of $\alpha, \tau_d$ and observe the variance for mean $q_0$. In \ref{qselect_app}, we display examples of how this wavenumber selection manifests, with the upper panel corresponding to lowest variance with and without noise, and the lower panel corresponds to the largest variance in all the parameter combinations. 
\subsection{Generalizations of the model}\label{2td}
As mentioned in the main, we extend our model by considering multiple memory kernels. In the first example, 
\begin{align}
    \partial_t \phi(\boldsymbol{r}, t) = \nabla^2 \mu_{eq} &+ \alpha_1 \nabla^2 \phi(\boldsymbol{r}, t- \tau_{d, 1}) \notag \\
    &+ \alpha_2 \nabla^2 \phi(\boldsymbol{r}, t- \tau_{d, 2}) \label{2td_model1}
\end{align}
and in the second example we consider a a different free energy functional for the second time delay, giving us:
\begin{align}
    \partial_t \phi(\boldsymbol{r}, t) = \nabla^2 \mu_{eq} &+ \alpha_1 \nabla^2 \phi(\boldsymbol{r}, t- \tau_{d, 1}) \notag \\
    &+ \alpha_2 \nabla^2 \left[ \phi(\boldsymbol{r}, t- \tau_{d, 2}) \right]^2\label{2td_model2}
\end{align}
In both the examples, when the feedback from the larger time delay $\tau_d^2 > \tau_d^1$ is tuned to be stronger, i.e $\alpha_2 > \alpha_1$, we see the emergence of warped limit cycles. But it looks different in each model. In \eqref{2td_model1}, we see non-uniform warped limit cycles at defects. But  in \eqref{2td_model2}, the dynamics looks completely different and the limit cycle is chaotic everywhere. When the feedback of smaller time delay is tuned to be stronger, i.e $\alpha_1 > \alpha_2$, both the examples show uniform limit cycles. 
Additionally, we perform linear stability analysis by perturbing a homogenous state to get the following equation:
\begin{align}
    i\omega &= -{q}^2 (3\bar{\phi}^2 - 1)  - \kappa {q}^4  \notag \\
    &- \alpha_1  {q}^2 e^{-i \omega \tau_d^1} - \alpha_2 q^2  e^{-i \omega \tau_d^2}
\end{align} 
The Eigenmodes are solved for numerically, using Newton-Raphson method and plotted in \ref{LS_2td}. We observe that unlike the single time delay case, the imaginary branches intersect with each other. 
\begin{figure}[h]
    \centering
    \includegraphics[width=0.7\linewidth]{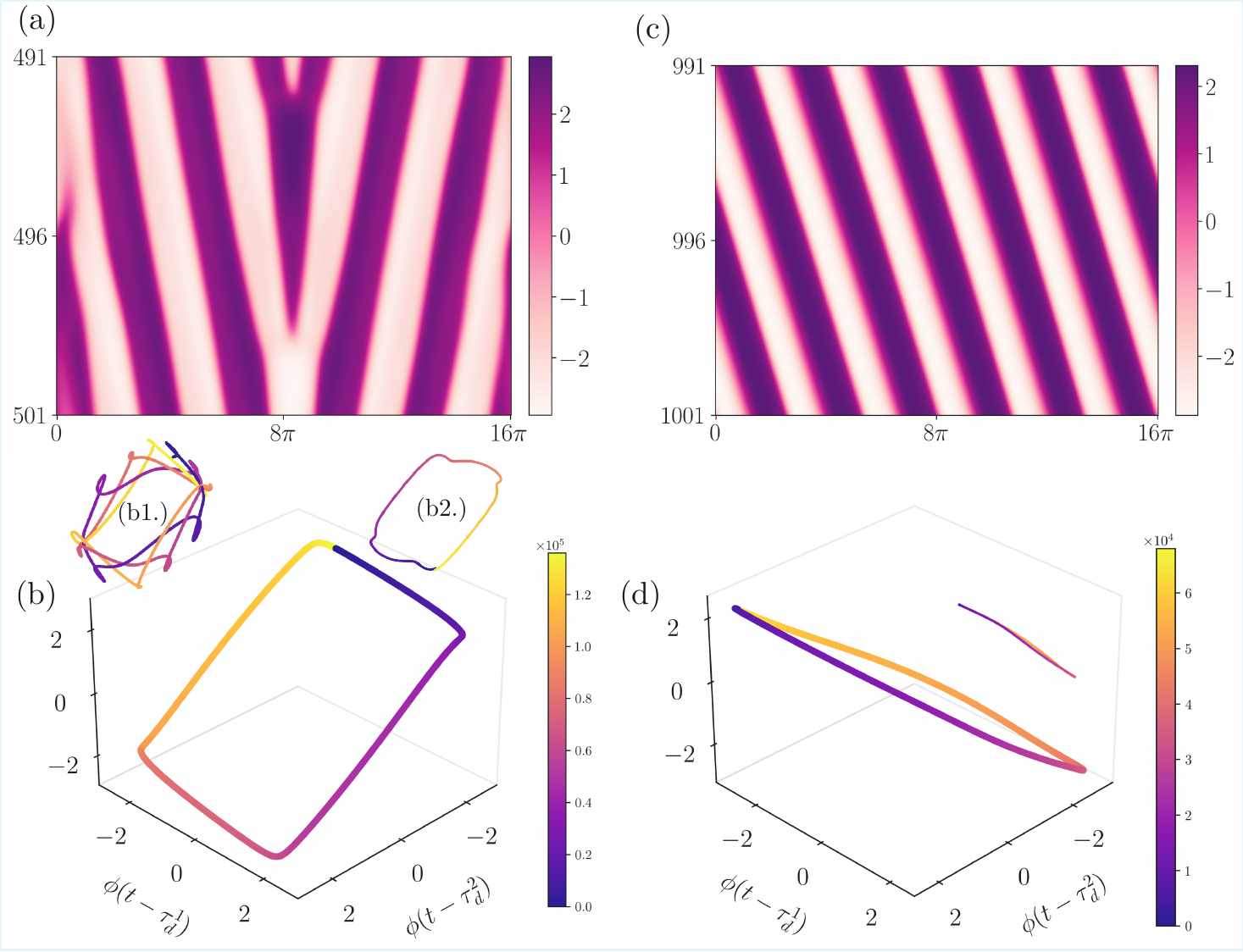}
    \caption{\textbf{Generalization to other forms} Panel (a), (b) shows the khymograph and limit cycles of \eqref{2td_model1} for $\alpha_1 = 2, \tau_{d, 1} = 3, \alpha_2 = 5, \tau_{d, 2} = 7$ respectively. The two subplots show limit cycles at different positions (b1) $x_0 = 0.4\pi$, while for (b2) at $x_0 = 8\pi$, where the defects are located. Here we observe that as opposed to the case presented in the main, the topology of defects are warped and non-uniform. Panels (c) and (d) display the kymograph and limit cycle of \eqref{2td_model2} for $\alpha_1 = 5$, $\tau_{d1} = 3$, $\alpha_2 = 2$, and $\tau_{d2} = 7$. In contrast to the chaotic limit cycles observed in the main text when $\alpha_2 > \alpha_1$, the limit cycle here is uniform. }
    \label{fig:enter-label}
\end{figure}
\begin{figure}
    \centering
    \includegraphics[width=0.8\linewidth]{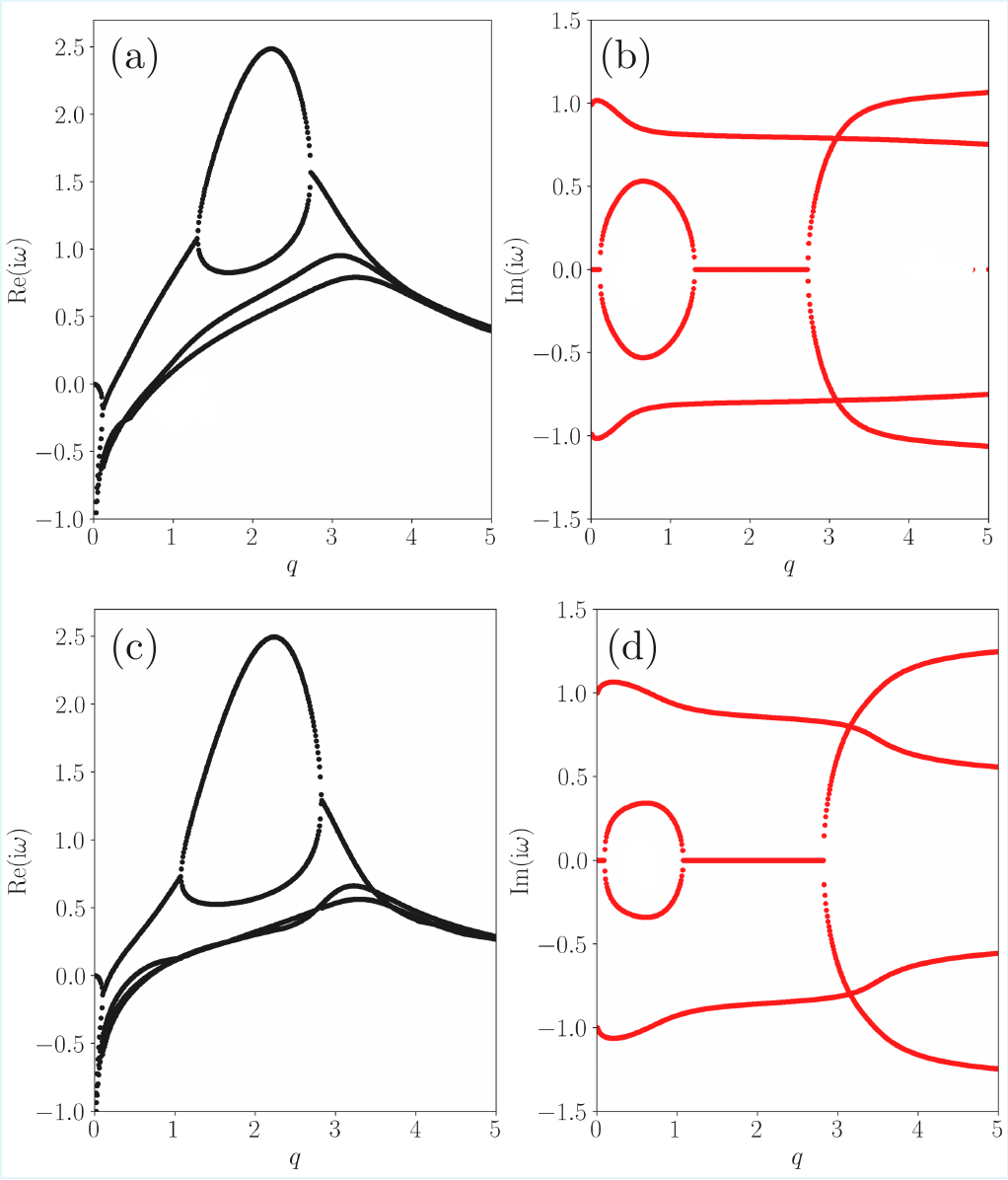}
    \caption{\textbf{Linear stability analysis for multiple time delay}Panels (a) - (b) for $\alpha_1 < \alpha_2$ and Panels (c) - (d) for $\alpha_2 > \alpha_1$, when $\tau_{d2} >\tau_{d1} $.}
    \label{LS_2td}
\end{figure}

\end{document}